\documentclass[letterpaper]{article}
\usepackage[totalwidth=480pt, totalheight=680pt]{geometry}

\usepackage[T1]{fontenc}
\usepackage{amsfonts,amsmath,amssymb,amsthm}
\usepackage{algorithm}
\usepackage{algorithmicx}
\usepackage{algpseudocode}
\usepackage{booktabs}
\usepackage{tabularx}
\usepackage{multirow}
\usepackage{placeins}
\usepackage{url}
\usepackage[round]{natbib}
\usepackage{xcolor}
\definecolor{darkgreen}{rgb}{0,0.5,0}
\usepackage{hyperref}
\hypersetup{
    colorlinks=true,
    linkcolor=red,
    citecolor=darkgreen,
    filecolor=magenta,
    urlcolor=blue
}
\usepackage{authblk}

\newtheorem{theorem}{Theorem}[section]
\newtheorem{proposition}[theorem]{Proposition}
\newtheorem{lemma}[theorem]{Lemma}
\newtheorem{corollary}[theorem]{Corollary}
\theoremstyle{definition}
\newtheorem{definition}[theorem]{Definition}

\theoremstyle{remark}

\makeatletter
\providecommand*{\theHALG@line}{}
\renewcommand*{\theHALG@line}{\thealgorithm.\arabic{ALG@line}}
\makeatother

\newcommand{\partprob}{\textsc{Partition}}

\title{\bf Closing Gaps in Temporal Fair Division for Chores}
\author[1]{Nicholas Teh}
\affil[1]{University of Oxford, UK}
\date{}

\begin{document}

\maketitle

\begin{abstract}
We study the fair division of indivisible chores that must be assigned in a prescribed order. The complete instance is known in advance, but assignments cannot be changed. Temporal envy-freeness up to one chore (TEF1) requires envy-freeness up to one chore after every arrival. For goods, TEF1 allocations need not exist, and deciding whether one exists is NP-hard. For chores, Elkind et al.\ (AAMAS 2025) showed that TEF1 allocations exist for two agents, but left open whether they always exist, conjecturing that they do not, and whether deciding their existence is hard. We resolve both questions, confirming the conjecture in a strong form: for every $n\ge3$ and every finite factor, some $n$-agent instance admits no allocation satisfying TEF1 even up to that factor. This holds even when all agents have the same multiset of individual chore costs, with nine chores sufficing for three agents. Deciding TEF1 existence is NP-complete for every fixed $n\ge3$, even when all costs lie within a factor $1+2^{-m}$ of one another, where $m$ is the number of chores. Instances without TEF1 allocations exist arbitrarily close to the uniform cost matrix with $O(n\log n)$ chores, and just nineteen for three agents.

We then relate fairness to social cost. Even when a TEF1 allocation is supplied, minimizing social cost subject to TEF1 admits no polynomial-time approximation within any factor polynomial in the number of chores unless P$=$NP, already for three agents with strictly positive normalized costs. In contrast, for every fixed number of agents, we give an FPTAS for the optimal temporal EF1 factor, the smallest factor up to which TEF1 can be satisfied, whose output has social cost no larger than that of the cheapest allocation attaining this factor. Thus a slight relaxation of fairness permits a social-cost guarantee that, unless P$=$NP, no polynomial-time algorithm can provide under exact TEF1.
\end{abstract}

\section{Introduction}

Dividing chores (i.e., tasks that impose a cost on whoever performs them) among a
group of agents is a fundamental problem in fair division, with applications
ranging from splitting household tasks among roommates to distributing teaching
loads among faculty members~\citep{AzizEtAl2022,IgarashiYokoyama2023}. In many
such applications, the chores are carried out over time.
A team may divide maintenance jobs, on-call duties, or other recurring tasks
fairly in the long run while leaving one member with a disproportionate burden
along the way. Fairness at the end is then not enough: the assignment should
also remain fair while the tasks are carried out. This matters even when all
tasks are known beforehand, since a duty that has already been performed cannot
later be transferred to someone else.

A prominent fairness notion is \emph{envy-freeness}
\citep{Foley1967,Varian1974}, which requires
every agent to find her own bundle no more costly than any other agent's
bundle. For indivisible tasks, however, envy-freeness is often too demanding:
if a single unpleasant chore is to be assigned to one of two agents, the agent
who receives it will envy the other. A standard relaxation is
\emph{envy-freeness up to one item} (EF1)
\citep{LiptonEtAl2004,Budish2011}, which was extended to chores by \citet{AzizEtAl2022JAAMAS}. For chores, it asks that every agent weakly prefer
her own bundle to any other agent's bundle after removing at most one of her
own chores, where both bundles are evaluated according to her costs.

When chores arrive over time,
\emph{temporal EF1} (TEF1) imposes this requirement after every arrival,
rather than only when all chores have been assigned. Its multiplicative
relaxation~\citep[cf.][]{AmanatidisBirmpasMarkakis2018,ZhouWu2024},
$\alpha$-TEF1 for a factor $\alpha\ge1$, requires only that an agent's
remaining cost be at most $\alpha$ times her cost for any other agent's
bundle; the smallest such $\alpha$ is the \emph{temporal EF1 factor} of the
allocation. We study the
informed model~\citep{HeEtAl2019,ElkindEtAl2025}: the costs and arrival order are known in advance, but every
chore must be assigned upon arrival and earlier assignments cannot be changed.

Since earlier assignments cannot be revised, knowing the future does not
reduce the problem to a sequence of static ones.
One might try to choose an EF1 allocation separately for each prefix of the
arrival sequence, but these allocations need not agree on the recipients of
chores that have already arrived. Alternatively, one might extend the current
allocation at each arrival while keeping it EF1, but an extension that is fair
now can leave no fair choice later. Neither approach settles whether a TEF1
allocation always exists.

A three-chore example illustrates the distinction. Suppose the costs, in
arrival order, are $(3,1,3)$ for the first agent and $(1,3,3)$ for the second.
Give the first chore to the first agent and the second chore to the second.
Each bundle contains at most one chore, so the first two prefixes are EF1. Yet
neither agent can receive the third chore: even after removing a largest own chore, its recipient has
cost $3$ left and evaluates the other bundle at only $1$. Giving the first
two chores to the opposite recipients avoids this difficulty and allows
either agent to receive the last chore. Thus foresight can help choose the
initial allocation, but an arbitrary fair initial choice need not extend.
Whether foresight is always enough (that is, whether a TEF1 allocation
always exists) is a separate question.

\citet{ElkindEtAl2025} studied TEF1 in this informed model for
both goods and chores. They showed that TEF1 allocations always exist and can be
computed efficiently for two agents and for several restricted preference
classes. For goods, TEF1 allocations need not exist in general (an instance
of \citet{HeEtAl2019} with three agents and $23$ goods admits
none) and Elkind et al.\ proved that deciding their existence is NP-hard.
For chores, however, they observed that the construction of He et al.\ does
not carry over. They established hardness of completing a given partial
allocation of chores, but left open whether TEF1 allocations always
exist (conjecturing that they do not) and whether deciding their
existence, without a given partial allocation, is computationally hard.

In this paper, we resolve both questions: we show that TEF1 allocations of
chores need not exist, thereby confirming the conjecture of Elkind et al.,
and that deciding whether they exist is NP-complete for every fixed number of at least three agents. Besides fairness, a
natural concern is efficiency: a chore that one agent can complete quickly may
take another agent much longer, so it is also desirable to keep the
\emph{social cost}, the total cost incurred by the agents, low.
If exact fairness cannot always be maintained, how much must we relax it, and can we efficiently find
an allocation that is nearly as fair as possible without incurring
unnecessary social cost?

\subsection{Our Results}

Our results distinguish between a fairness guarantee that holds for all
instances and an approximation to the optimal temporal EF1 factor of each
instance. No finite guarantee of the first
kind exists even with three agents, whereas the optimal factor admits an
approximation scheme for every fixed number of agents. We also show why
relaxing fairness matters for social cost.

\paragraph{Unbounded temporal EF1 factors with at least three agents.}
Section~\ref{sec:nonexistence} gives instances for every $n\ge3$ with
$n+O(\log n)$ chores for which the optimal temporal EF1 factor is unbounded;
nine chores suffice for $n=3$. Thus
no fixed factor, however large, can be guaranteed; in particular, TEF1
allocations need not exist. The number of chores does not grow with
the desired factor, and every agent has the same multiset of individual costs.
Hence the unbounded factor does not stem from agents evaluating chores on
different scales. Combined with the existence of TEF1 allocations for two
agents~\citep{ElkindEtAl2025}, our result shows a qualitative change from two
to three agents: with three agents, not even a prescribed multiplicative
relaxation of TEF1 can always be satisfied.

\paragraph{Nearly equal costs do not make exact fairness easy.}
Section~\ref{sec:uniform} proves that deciding TEF1 existence is NP-complete
for every fixed $n\ge3$, even when the largest entry of the entire cost matrix is at
most $1+2^{-m}$ times the smallest, where $m$ is the number of chores. Hence
the hardness does not rely on large differences between chore costs. We also
give $n$-agent instances with $O(n\log n)$ chores that admit no TEF1 allocation
and, after scaling, converge to the all-ones cost matrix, which itself admits
a TEF1 allocation. These
results show that exact TEF1 is sensitive to small changes in costs, even
though a temporal EF1 factor close to $1$ is easy to achieve when costs are
nearly uniform.

\paragraph{Exact TEF1 can be costly, and its minimum cost is hard to approximate.}
Section~\ref{sec:cost} first gives a two-agent example in which requiring TEF1
increases the minimum social cost by an arbitrarily large factor, although
an arbitrarily small multiplicative relaxation attains the unconstrained
minimum. Some allocation attaining the unconstrained minimum is EF1 at the
final round, so the cost increase is due specifically to fairness at earlier
rounds. In the static setting, by contrast, requiring EF1
with two agents increases the minimum social cost by a factor of at most
$5/4$~\citep{SunChenDoan2023}.
We then prove that, unless P$=$NP, no polynomial-time algorithm can
approximate the minimum social cost of a TEF1 allocation within any factor
polynomial in the number of chores. This holds for any fixed number of at
least three agents, even with strictly positive normalized costs, and even
when the input includes a TEF1 allocation.

\paragraph{Near-optimal temporal EF1 factor without social-cost loss.}
Section~\ref{sec:approx} gives a fully polynomial time approximation scheme
(FPTAS) for the optimal temporal EF1 factor when the number of agents
is fixed. The output also has social cost no larger than the minimum among
allocations attaining the optimal temporal EF1 factor. More generally, for any
attainable target factor and any accuracy $\varepsilon>0$, the algorithm loses
only a factor $1+\varepsilon$ in fairness and loses nothing in social cost
relative to the cheapest allocation meeting the target. In particular, when
a TEF1 allocation exists, the algorithm returns a $(1+\varepsilon)$-TEF1
allocation whose social cost is at most the minimum social cost of a TEF1
allocation (a benchmark that, unless P$=$NP, no polynomial-time algorithm
can approximate within a polynomial factor while keeping TEF1 exact).

Table~\ref{tab:results-summary} summarizes our results.
Full proofs are in the appendix; the main text explains the
ideas behind our constructions and discusses their implications.

\begin{table}[t]
\caption{Summary of our results. Here, $n$ and $m$ denote the numbers of
agents and chores, $\alpha^*$ denotes the optimal temporal EF1 factor, and
$\mathrm{sc}^*$ denotes the minimum social cost among allocations
attaining $\alpha^*$. The thresholds satisfy $m_n=n+O(\log n)$,
$M_n=O(n\log n)$, $m_3=9$, and $M_3=19$.}
\label{tab:results-summary}
\centering
\small
\renewcommand{\arraystretch}{1.15}
\begin{tabularx}{\textwidth}{@{}>{\raggedright\arraybackslash}Xccl@{}}
\toprule
Result & $n$ & $m$ & Ref.\\
\midrule
\multicolumn{4}{@{}l}{\emph{Existence of fair allocations}}\\
For every $\alpha\ge1$, some instance admits no $\alpha$-TEF1 allocation
& $\ge3$ & $\ge m_n$ & Thm.~\ref{thm:unbounded}\\
Deciding TEF1 existence is NP-complete, even when all costs lie within a
factor $1+2^{-m}$ of one another
& fixed $\ge3$ & any & Thm.~\ref{thm:near-uniform}\\
Some instances with costs arbitrarily close to the all-ones matrix (after
scaling) admit no TEF1 allocation
& $\ge3$ & $\ge M_n$ & Cor.~\ref{cor:uniform-limit}\\
\midrule
\multicolumn{4}{@{}l}{\emph{Social cost}}\\
TEF1 can increase the minimum social cost unboundedly, even when
$(1+\varepsilon)$-TEF1 does not increase it
& $\ge2$ & $\ge n+3$ & Prop.~\ref{prop:cost-separation}\\
Minimum-cost TEF1 is inapproximable within $m^{O(1)}$ unless P$=$NP, even
given a TEF1 allocation
& $\ge3$ & any & Thm.~\ref{thm:cost-hardness}\\
An FPTAS for $\alpha^*$ returns an allocation with social cost at most
$\mathrm{sc}^*$
& fixed & any & Thm.~\ref{thm:fptas}\\
\bottomrule
\end{tabularx}
\end{table}

\subsection{Related Work}
\label{sec:related}

\paragraph{Temporal fair division.}
Our model is the informed setting of \citet{HeEtAl2019}, who showed
that, for goods, informed algorithms can maintain EF1 after every arrival
without reassignments for two agents but not for three.
\citet{ChoiLiTeh2026} recently identified further tractable settings
for mixed goods and chores; for instance, a cyclic rule is TEF1 whenever there
are at most two item types. Their dynamic programs run in polynomial
time when both the number of agents and the number of distinct item-value
vectors are fixed, and in pseudo-polynomial time for a fixed number of agents
with integer values. These results do not give running time
polynomial in the binary input length for arbitrary costs. By approximating
the optimal temporal EF1 factor instead, we obtain polynomial running
time for arbitrary chore costs and a fixed number of agents, while never
exceeding the social cost of the cheapest allocation attaining the optimal
factor. \citet{GoldbergRobinsonTeh2026} studied the informed
model for goods with a different objective, the maximum envy summed over all
rounds, and gave an FPTAS for a fixed number of agents with identical
valuations; our FPTAS allows arbitrary costs. \citet{AmanatidisEtAl2026Buffers} showed that bounded reordering of
arrivals yields EF1 after every step when each agent has few distinct values;
without reordering, our instance in Table~\ref{tab:nine-chores}, with four
distinct costs per agent, admits no TEF1 allocation despite complete foresight.

\paragraph{Approximating fairness.}
For static allocations of goods, \citet{LiptonEtAl2004}
introduced the problem of minimizing the envy ratio, and \citet{NguyenRothe2014} gave an FPTAS for it when the number of agents is
fixed. Our FPTAS groups the states of a dynamic program by geometric
intervals, a classical technique going back to \citet{IbarraKim1975} and used, for example, for scheduling on a fixed
number of unrelated machines~\citep{HorowitzSahni1976}; see
\citet{Woeginger2000} for general conditions under which this
approach applies. Our
objective differs from the envy ratio in two ways: it removes a largest own
chore before comparing costs, and it is evaluated at every prefix. Because of
the first, a small
error in a bundle's total cost need not give a small relative error after the
largest chore is removed; because of the second, good final fairness does not
determine fairness at earlier rounds. Handling these two differences is the
main difficulty in our algorithm.

\paragraph{Fairness and efficiency.}
\citet{ElkindEtAl2025} showed that TEF1 can be incompatible with
Pareto optimality, which requires that no reassignment reduce one agent's
cost without increasing another's, and that deciding whether both can be
attained is NP-hard. In the static setting, allocations of chores that are
weighted EF1 and fractionally Pareto optimal can be computed in polynomial
time when the number of chores exceeds the number of agents by at most a
fixed constant~\citep{Teh2026Surplus}. Our
inapproximability result for social cost compares an allocation with the
cheapest TEF1 allocation, rather than with an unconstrained efficient
allocation. In the static chores setting, \citet{SunChenDoan2023}
showed that, for normalized costs, the ratio between the minimum social cost
of EF1 allocations and the unconstrained minimum social cost is at most $5/4$
for two agents, but can be arbitrarily large for three or more agents.
With mixed goods and chores, by contrast, the utilitarian price of EF1 is
unbounded already for two agents~\citep{Teh2026MixedManna}.
\citet{LiSunXing2024} proved that, unless P$=$NP, minimum-cost EF1
admits no bounded multiplicative approximation for four or more agents.
For goods, the maximum welfare of an EF1 allocation is hard to approximate
within polynomial factors when the number of agents is part of the
input~\citep{BarmanEtAl2019,BuEtAl2025}, but can be approximated within a
constant factor for every constant number of agents~\citep{BuEtAl2025}, and
\citet{LimNeohTeh2026Cost} studied generalized-mean welfare within EFX
allocations when the goods outnumber the agents by at most three.
For chores with temporal fairness, the corresponding cost ratio is unbounded
already for two agents, and
we establish inapproximability within every polynomial factor for any fixed
number of at least three agents,
even when a TEF1 allocation is supplied. We complement these results with an
algorithm that slightly relaxes fairness but loses nothing in social cost;
for goods and a constant number of agents, \citet{BuEtAl2025} showed
that a $(1-\varepsilon)$-relaxation of static EF1 similarly permits optimal
welfare. Unlike theirs, our relaxation is relative to an optimal factor
that may be arbitrarily large, and it must hold at every prefix.

\paragraph{Nonexistence of fair allocations.}
Maximin share allocations can fail to exist already with three agents and
nine goods or nine chores~\citep{FeigeSapirTauber2021}, and for three
agents with superadditive chore costs, no multiplicative approximation of EFX
can be guaranteed~\citep{ChristoforidisSantorinaios2024}. For two agents with
strictly increasing valuations over goods, EF1 and Pareto optimality can be
incompatible, and eight goods are necessary and sufficient for such an
instance~\citep{Teh2026Increasing}. For additive chores,
EF1 allocations always
exist~\citep{AzizEtAl2022JAAMAS}, and so do constant-factor approximations of
EFX~\citep{GargMurhekarQin2025}; EFX allocations of chores also exist under
budget constraints~\citep{ElkindIgarashiTeh2024}, and weighted EF1
allocations exist even for mixed goods and chores~\citep{Teh2026MixedManna}.
Theorem~\ref{thm:unbounded}
shows that, at every prefix, not even EF1 admits a multiplicative guarantee,
already for three agents with additive costs.

\paragraph{Other models.}
Other temporal models allow changes to earlier assignments
\citep{HeEtAl2019,HuangTeh2026Chores,HuangTeh2026Goods}, impose fairness both within and across days
\citep{CooksonEtAl2025}, or repeatedly allocate the same collection of items
\citep{IgarashiEtAl2024,LimNeohTeh2026}; \citet{ChoiLi2026a,ChoiLi2026b} studied
temporal versions of EFX and maximin share fairness for goods, in part with
the option of delaying items. Under upgrade recourse, where each chore
previously assigned to an agent may be replaced only by a distinct chore that
she finds no more costly, a simple adaptation of round robin maintains EF1
without any information about future chores~\citep{HuangTeh2026Chores}. By
Theorem~\ref{thm:unbounded}, complete information about the future cannot
make up for the absence of such revisions. Fairness over time has also been
studied in voting~\citep{Lackner2020,ElkindObraztsovaTeh2024}, including for public
chores~\citep{ElkindNeohTeh2025}. Online fair division instead limits
knowledge of future
arrivals~\citep{AleksandrovEtAl2015,AleksandrovWalsh2020}; in this model,
\citet{BenadeEtAl2024OR} studied the growth of envy over time
and its trade-offs with efficiency, fairness guarantees for chores have been
investigated~\citep{ZhouBaiWu2023,SongEtAl2026}, and \citet{NeohTeh2026} recently obtained impossibility results for several
fairness notions. Predictions about future items can restore share-based
guarantees~\citep{NeohPetersTeh2026,ChooEtAl2026}, although knowing each
agent's multiset of values in advance does not permit any positive
approximation of EF1 for five or more agents~\citep{NeohTeh2026Frequency}. Our
nonexistence results allow complete knowledge of the future and hence do not
rely on uncertainty about later chores. Appendix~\ref{app:related} discusses
further related work.

\section{Model and Preliminaries}
\label{sec:model}

For a positive integer $k$, write $[k]=\{1,\dots,k\}$. Let $N=[n]$ be
the set of agents, where $n\ge2$, and let $C=(c_1,\dots,c_m)$ be an
ordered list of chores. Chore $c_t$ arrives in round
$t$ and must be assigned in that round. Each agent $i$ has a nonnegative
additive cost function $d_i$; thus $d_i(S)=\sum_{c\in S}d_i(c)$ for every set
$S$ of chores. The instance and its arrival order are known in advance.
The complexity results use the standard binary encoding of costs. Costs
are \emph{normalized} if $d_i(C)=1$ for every agent $i$.

An allocation is a map $\pi:[m]\to N$. It specifies the recipient of each
chore and thereby determines a partial allocation after every round. Let $A_i^t=\{c_s:s\le t,\ \pi(s)=i\}$ and $A^t=(A_1^t,\dots,A_n^t)$.
Initially, $A_i^0=\varnothing$ for every agent $i$. We also call $A^t$ the
\emph{prefix} of $\pi$ at round $t$. Fairness will be required of every
prefix, not only of the final allocation $A^m$.

For a partial allocation $A$, let $M_i(A)=\max(\{d_i(c):c\in A_i\}\cup\{0\})$ and $R_i(A)=d_i(A_i)-M_i(A)$.
We call $R_i(A)$ agent $i$'s \emph{residual cost}. It is the least cost that
can remain after removing at most one chore from her own bundle. In
particular, the residual is zero for an empty bundle or a singleton.
This gives a convenient equivalent definition of EF1.

\begin{definition}[EF1 and TEF1]
A partial allocation $A$ is \emph{envy-free up to one chore} (EF1) if $R_i(A)\le d_i(A_j)$ for all distinct $i,j\in N$.
An allocation $\pi$ is \emph{temporal EF1} (TEF1) if $A^t$ is EF1 for every
$t\in[m]$.
\end{definition}

\begin{definition}[$\alpha$-TEF1]
For $\alpha\ge1$, an allocation $\pi$ is \emph{$\alpha$-TEF1} if $R_i(A^t)\le\alpha\,d_i(A_j^t)$ for all $t\in[m]$ and distinct $i,j \in N$.
\end{definition}

Thus a smaller factor is better, and $1$-TEF1 is exactly TEF1. The
\emph{temporal EF1 factor} of an allocation $\pi$ and the \emph{optimal
temporal EF1 factor} of the instance are
\begin{equation}\label{eq:factor}
 \alpha(\pi)=\max\left\{1,\max_{t,\,i\ne j}
 \frac{R_i(A^t)}{d_i(A_j^t)}\right\},
 \quad \alpha^*=\min_\pi\alpha(\pi),
\end{equation}
where $0/0=0$ and $x/0=+\infty$ for $x>0$.
Because the outer maximum in \eqref{eq:factor} includes $1$, using $0/0=1$
instead would give the same value of $\alpha(\pi)$.
The convention $x/0=+\infty$ reflects that no finite factor permits a positive
residual when the other bundle is empty or has zero cost.
For example, in the three-chore instance from the introduction, every
allocation that gives the first chore to the first agent and the second chore
to the second agent has temporal EF1 factor $3$, whereas every allocation that
gives the first two chores to the opposite recipients has factor $1$.

\begin{lemma}\label{lem:finite-factor}
Every instance admits an allocation with finite temporal EF1 factor.
\end{lemma}
\begin{proof}
Call a chore fully positive if every agent assigns it positive cost. Assign
each chore that is not fully positive to an agent who assigns it zero cost.
Assign the fully positive chores, in their arrival order, cyclically among the
agents.

Consider any prefix $A^t$ and any agent $i$ with $R_i(A^t)>0$. Only fully
positive chores can contribute positive cost to $i$'s own bundle, and a
positive residual means that $i$ has received at least two such chores. By the
cyclic assignment, every other agent has then received at least one fully
positive chore. Hence $d_i(A_j^t)>0$ for every $j\ne i$. Thus every ratio in
\eqref{eq:factor} with positive numerator has positive denominator, so this
allocation has finite factor.
\end{proof}

Thus $\alpha^*$ is finite on every instance, so approximating the optimal
temporal EF1 factor is always meaningful.

The \emph{social cost} of an allocation is the total cost incurred by its
recipients: $\mathrm{sc}(\pi)=\sum_{t=1}^m d_{\pi(t)}(c_t)$.
As a benchmark, we first minimize the temporal EF1 factor and then minimize
social cost among allocations attaining that factor; that is, let
$\mathrm{sc}^*=\min\{\mathrm{sc}(\pi):\alpha(\pi)=\alpha^*\}$.
When a TEF1 allocation exists, $\alpha^*=1$ and $\mathrm{sc}^*$ is the
minimum social cost of a TEF1 allocation. Otherwise, $\mathrm{sc}^*$
is still well defined, since it is taken over allocations attaining
$\alpha^*$ rather than over TEF1 allocations.

Two simple facts will be useful. When a chore is assigned, only its
recipient's ratios in \eqref{eq:factor} can increase: every other agent's own
bundle stays fixed, and her comparison bundles, namely the other agents'
bundles, can only become more costly.
Also, if chore $c_t$ is given to agent $i$, her residual becomes
\begin{equation}\label{eq:residual-update}
 R_i(A^t)=R_i(A^{t-1})+\min\{M_i(A^{t-1}),d_i(c_t)\}.
\end{equation}
Indeed, her own bundle now costs
$R_i(A^{t-1})+M_i(A^{t-1})+d_i(c_t)$, and the new maximum is $\max\{M_i(A^{t-1}),d_i(c_t)\}$. The nonnegative increment in
\eqref{eq:residual-update} will be important for approximating the optimal
factor.

\section{Unbounded Temporal EF1 Factors}
\label{sec:nonexistence}

The existence of TEF1 allocations for two agents~\citep{ElkindEtAl2025} might
suggest that an additional agent only requires a somewhat larger fairness
factor. Our first result rules out this possibility: the required factor can
grow without bound even when both the number of agents and the number of
chores remain fixed.

\begin{theorem}\label{thm:unbounded}
There are integers $m_n=n+O(\log n)$, $n\ge3$, with $m_3=9$, such that,
for every $\alpha\ge1$ and $m\ge m_n$, some $n$-agent, $m$-chore instance
with strictly positive costs admits no $\alpha$-TEF1 allocation. All agents
can have the same multiset of individual chore costs. The claim also holds
with normalized costs.
\end{theorem}

For $n=3$, label the agents $p,d,q$. Table~\ref{tab:powers} gives the construction
for an integer $h\ge2$; its optimal factor satisfies $\alpha^*>h-1$.
Each row is a permutation of $1,h,h^2,\dots,h^6,h^7,h^7$.
In particular, increasing $h$ changes the relative importance of chores,
not the total cost of one agent relative to another. The last two chores
have the same cost for all three agents.

\begin{table}[t]
\caption{For $h\ge2$, the optimal temporal EF1 factor is greater than $h-1$.
The arrival order is from left to right.}
\label{tab:powers}
\centering
\begin{tabular}{c|ccccccccc}
\toprule
 & $c_1$ & $c_2$ & $c_3$ & $c_4$ & $c_5$ & $c_6$ & $c_7$ & $c_8$ & $c_9$\\
\midrule
$p$ & $h^2$ & $1$ & $h$ & $h^3$ & $h^5$ & $h^6$ & $h^4$ & $h^7$ & $h^7$\\
$d$ & $h$ & $h^2$ & $1$ & $h^3$ & $h^4$ & $h^6$ & $h^5$ & $h^7$ & $h^7$\\
$q$ & $h^3$ & $h$ & $1$ & $h^2$ & $h^5$ & $h^4$ & $h^6$ & $h^7$ & $h^7$\\
\bottomrule
\end{tabular}
\end{table}

To see why these costs cause difficulty, note that once an agent owns two
sufficiently costly chores, removing the more costly one still leaves a chore that can
outweigh another agent's entire bundle. Geometric costs make this comparison
particularly transparent: a chore of cost $h^k$ is more than $h-1$ times the
sum of all smaller powers appearing in that row. Consequently, in an
$(h-1)$-TEF1 allocation, an agent's second-largest own cost can never exceed
the largest cost, according to her, in another agent's bundle. Since all
costs are positive, the first three chores must go to distinct agents, and
the arrival order then makes it impossible to maintain all of these
comparisons through the ninth chore. Appendix~\ref{app:nonexistence} gives
the complete calculation. For $n>3$, $n-3$ initial chores restrict later
assignments to three agents, and $O(\log n)$ nine-chore blocks include the
three rows for every possible remaining triple.

Theorem~\ref{thm:unbounded} does not contradict the fact that, for each $t$,
the first $t$ chores admit an EF1 allocation~\citep{AzizEtAl2022JAAMAS}. Rather, it
shows that such allocations for different values of $t$ cannot always be
chosen so that each extends the previous one. In particular,
requiring fairness only at the final round would remove the difficulty in
Theorem~\ref{thm:unbounded}.

The theorem rules out the prescribed factor regardless of the allocation
rule or computation time. For three agents, nine chores suffice for every
factor, whereas the instance of \citet{HeEtAl2019} that rules out
exact TEF1 for goods has $23$ goods. Appending chores preserves nonexistence,
since fairness is required after every round. Hence no finite bound depending
only on the numbers of agents and chores can hold in general, and the
guarantee in Section~\ref{sec:approx} must be relative to $\alpha^*$.

Ruling out exact TEF1 does not require large costs.
Table~\ref{tab:nine-chores} gives a second three-agent, nine-chore
instance, with positive integer costs at most $10$, that admits no TEF1
allocation; Appendix~\ref{app:small-instance} verifies this by checking every
possible partial allocation. Each agent has only four distinct costs. This
smaller numerical example will also be useful for the computational results:
changing the cost vector of its last chore yields an instance with a unique
TEF1 allocation, from which both computational reductions can begin.

\begin{table}[t]
\caption{A nine-chore instance with costs at most $10$ that admits no TEF1
allocation. The arrival order is from left to right.}
\label{tab:nine-chores}
\centering
\begin{tabular}{c|rrrrrrrrr}
\toprule
 & $c_1$ & $c_2$ & $c_3$ & $c_4$ & $c_5$ & $c_6$ & $c_7$ & $c_8$ & $c_9$ \\
\midrule
$p$ & $2$ & $1$ & $2$ & $2$ & $8$ & $8$ & $4$ & $8$ & $8$ \\
$d$ & $2$ & $3$ & $1$ & $3$ & $3$ & $5$ & $5$ & $5$ & $5$ \\
$q$ & $5$ & $2$ & $1$ & $2$ & $10$ & $5$ & $10$ & $10$ & $10$ \\
\bottomrule
\end{tabular}
\end{table}

\section{Exact Fairness with Nearly Equal Costs}
\label{sec:uniform}

The unbounded factors in Theorem~\ref{thm:unbounded} arise from large
differences between an agent's costs for different chores. We next ask
whether exact TEF1 becomes easy when these differences are small.
When all costs are equal, assigning the chores cyclically yields a TEF1
allocation: at each round, bundle sizes differ by at most one, so removing one
own chore is enough. This also follows from results for two item
types~\citep{ElkindEtAl2025,ChoiLiTeh2026}. This favorable situation does not
persist when costs are only slightly different, for either existence or
computation.

Let
\textsc{TEF1-Chores} denote the problem of deciding whether a given instance
admits a TEF1 allocation.

\begin{theorem}\label{thm:near-uniform}
\textsc{TEF1-Chores} is NP-complete for every fixed number $n\ge3$ of
agents with strictly positive integer costs, even when
$\max_{i,c}d_i(c)\le(1+2^{-m})\min_{i,c}d_i(c)$.
In particular, for every fixed $\eta>0$, NP-completeness holds when all
costs belong to $[K,(1+\eta)K]$ for some common positive integer $K$.
\end{theorem}

Thus the allowed relative difference between the largest and smallest costs
decreases exponentially with the number of chores. A single interval
contains the entire matrix, so costs are close across agents as well as
within each agent's row.
Since TEF1 existence can be decided in pseudo-polynomial time for a fixed
number of agents, and in polynomial time when the number of distinct cost
vectors is also fixed~\citep{ChoiLiTeh2026}, any NP-hardness proof for a
fixed number of agents must use large numbers and many distinct cost vectors
unless P$=$NP; Theorem~\ref{thm:near-uniform} shows that the large numbers
need not lead to large cost ratios.

The proof has two steps. First, we prove hardness of the existence question
itself, in which no partial allocation is given as part of the input.
For three agents, the constructed instance begins with the modified nine-chore instance mentioned
above, whose unique TEF1 allocation fixes the assignment of the first nine
chores. Later chores encode an instance of \partprob{}, which asks whether
a list of positive integers can be divided into two parts of equal sum.
Agents $p$ and $q$ have the same cost for each later chore, whereas agent
$d$'s costs for these chores are $1,2,4,\dots$, so each exceeds the sum of
all earlier ones. Since $d$'s EF1 inequality toward $p$ holds with equality
after the first nine chores, these costs prevent her from receiving any later
chore.
Comparing each of $p$ and $q$ with $d$'s unchanged bundle then bounds the
total cost of the later chores that each of them can receive, and the two bounds
add up exactly to the total cost of these chores. Both bounds must therefore
be attained, which is possible exactly when the \partprob{} instance has a
solution. Appendices~\ref{app:hardness} and~\ref{app:add-agents} give this
reduction and extend it to every fixed $n\ge3$ by adding initial chores that
force all later chores to go to the original three agents.

Second, we show how to make all costs nearly equal without changing whether
a TEF1 allocation exists. We add a sufficiently large common amount $K$ to
every original cost, which is a standard way of controlling
cardinalities~\citep{GareyJohnson1979}. To make this compatible with fairness
at every prefix, we also insert additional chores whose cost is exactly $K$
for every agent. In a TEF1 allocation of the new instance, bundle sizes can differ by at
most one: if an agent had two more chores than another agent, the resulting
extra cost of at least $K$ could not be offset by the much smaller original
costs. The inserted chores restore equal bundle sizes before each subsequent
original chore. When that chore is assigned, its recipient has exactly one
more chore than every other agent, so removing one own chore cancels the
common amounts $K$ and leaves the original EF1 inequality. The full
argument in Appendix~\ref{app:near-uniform} proves both that TEF1 allocations
of the new instance yield TEF1 allocations of the original one and that every
original TEF1 allocation extends through all inserted chores.

Applying the same transformation to our nonexistence instances keeps the
number of chores fixed as the costs approach uniformity.

\begin{corollary}\label{cor:uniform-limit}
For every $n\ge3$, there are positive integers $M_n=O(n\log n)$, $D_n$,
and $S_n$ such that, for every integer $K>S_n$ and $m\ge M_n$, some
$n$-agent, $m$-chore instance with costs in $[K,K+D_n]$ admits no TEF1
allocation. One may take $(M_3,D_3,S_3)=(19,10,55)$. For fixed $n,m$, after
dividing all costs by $K$, these instances converge to the all-ones cost
matrix, which admits a TEF1 allocation.
\end{corollary}

Unlike in Theorem~\ref{thm:unbounded}, a temporal EF1 factor close to $1$
is easy to obtain when costs are nearly uniform. If every cost lies in
$[K,(1+\eta)K]$, assigning the chores cyclically yields a
$(1+\eta)$-TEF1 allocation. Indeed, if agent $i$ owns $k_i$
chores and agent $j$ owns $k_j$, then $k_i-1\le k_j$ whenever $k_i>0$.
After one own chore is removed, the remaining cost is at most
$(k_i-1)(1+\eta)K$, while $j$'s bundle costs at least $k_jK$.
Together with this observation, Theorem~\ref{thm:near-uniform} shows that
the difficulty lies in requiring exact TEF1, even when a $(1+\eta)$-TEF1 allocation is
immediate.

\section{Social Cost and a Small Fairness Relaxation}
\label{sec:cost}

Existence is not the only reason to consider a relaxation. Even when TEF1
allocations exist, insisting on exact fairness at each round can force a
chore onto an agent who finds it much more costly than another agent does. We first show that an
arbitrarily small relaxation can eliminate an unbounded cost increase.
We then ask whether one can at least approximate the minimum social cost of a
TEF1 allocation while keeping fairness exact.

\subsection{The Cost of Fairness at Earlier Rounds}

For static allocations of chores with normalized costs, the worst-case ratio
between the minimum social cost of an EF1 allocation and the unconstrained
minimum social cost (the \emph{price of fairness}~\citep{BertsimasFariasTrichakis2011,CaragiannisEtAl2012}
with respect to EF1) is unbounded when there are at least three agents, but
only $5/4$ when there are two agents~\citep{SunChenDoan2023}. To distinguish this static
effect from the temporal requirement, our next construction makes an
unconstrained minimum-cost allocation EF1 at the final round, so that only an
inequality at an earlier round prevents it from being TEF1.

\begin{proposition}\label{prop:cost-separation}
For every $n\ge2$, $m\ge n+3$, and $0<\varepsilon\le1$, there are
$n$-agent, $m$-chore instances with strictly positive normalized costs for
which the ratio between the minimum social cost of a TEF1 allocation and the
unconstrained minimum social cost is arbitrarily large. Nevertheless, some $(1+\varepsilon)$-TEF1 allocation
attains the unconstrained minimum social cost and is EF1 at the final round.
\end{proposition}

Table~\ref{tab:cost-example} gives two-agent, five-chore instances, where
$H>2$. Both agents
have total cost $3H+1+\varepsilon$, so dividing all costs by this total
normalizes the instance. Consider the recipient sequence $1,2,1,1,2$.
It assigns each chore to a minimum-cost agent and has social cost
$3+2\varepsilon$ before normalization. After the fourth chore, agent $1$
can remove a chore of cost $1$, but her remaining cost is
$1+\varepsilon$ and her comparison bundle costs only $1$.
Only this inequality requires the factor $1+\varepsilon$. Giving the final
chore to agent $2$ restores exact EF1 at the final round.

\begin{table}[t]
\caption{An arbitrarily small relaxation of TEF1 can eliminate an unbounded
social-cost increase. Here $0<\varepsilon\le1$ and $H>2$; divide by
$3H+1+\varepsilon$ for normalized costs.}
\label{tab:cost-example}
\centering
\begin{tabular}{c|ccccc}
\toprule
 & $c_1$ & $c_2$ & $c_3$ & $c_4$ & $c_5$\\
\midrule
$1$ & $\varepsilon$ & $1$ & $1$ & $1$ & $3H-2$\\
$2$ & $H$ & $1$ & $H$ & $H$ & $\varepsilon$\\
\bottomrule
\end{tabular}
\end{table}

An exact TEF1 allocation cannot make the same inexpensive choices for the
first four chores. If chores $c_1,c_3,c_4$ all go to agent $1$, then $c_2$
must go to agent $2$, since with positive costs the first two chores cannot
have the same recipient; agent $1$'s EF1 inequality after the fourth chore
is then violated.
Hence at least one of these three chores must go to agent $2$, at cost $H$.
Exact TEF1 is still attainable; Appendix~\ref{app:cost-separation} checks
a complete such allocation. Letting $H$ grow proves the claimed cost ratio.
Appendix~\ref{app:cost-separation} also extends the construction to any
$n\ge2$ agents and $m\ge n+3$ chores.

\subsection{Optimizing Cost While Keeping Fairness}

The preceding example compares exact TEF1 with an unconstrained optimum.
A more modest computational objective is to compare only with other TEF1
allocations. Even if some cost increase is unavoidable, one might hope to
efficiently find a TEF1 allocation whose social cost is close to the minimum
among TEF1 allocations. Our next result rules out even an approximation ratio
that grows polynomially with the number of chores.

\begin{theorem}\label{thm:cost-hardness}
Unless P$=$NP, there is no polynomial-time $m^{O(1)}$-approximation algorithm
for minimizing social cost among TEF1 allocations for any fixed number
$n\ge3$ of agents, even with strictly positive normalized costs and even when
a TEF1 allocation is supplied with the instance.
\end{theorem}

Here $m^{O(1)}$ allows any fixed polynomial approximation ratio in the number
of chores, not just a constant. Supplying a TEF1 allocation with the
instance also strengthens the result. A hardness proof based on instances that may admit no TEF1 allocation would
not answer this optimization question, since an algorithm could be unable to
return an allocation simply because none exists. In our construction, every instance
has a TEF1 allocation that we provide explicitly; what is hard is finding
a cheap one.

The reduction uses the same nine-chore initial assignment as the existence
proof, but changes the later chores. Those chores are inexpensive for two
agents and expensive for the third. A TEF1 allocation can assign all of them
inexpensively exactly when the \partprob{} instance has a solution.
Additional pairs of chores ensure that, whether or not a solution exists, the
third agent can instead receive all chores that encode the \partprob{}
instance while TEF1 is maintained. Their number and costs are chosen so that these pairs do not
alter the partition encoded by a cheap allocation. The ratio between the
social costs of the expensive and cheap allocations can exceed any prescribed
polynomial in the number of chores. A final chore makes all row totals
equal, so that the costs can be normalized. Appendix~\ref{app:cost-hardness}
verifies the supplied allocation, the cost comparison, and the
normalization, and extends the construction to more than three agents.

Theorem~\ref{thm:cost-hardness} concerns the minimum social cost among TEF1
allocations, whereas Proposition~\ref{prop:cost-separation} concerns the
cost of imposing temporal fairness at all. Together they motivate allowing
a small fairness relaxation rather than demanding exact fairness and
approximating only cost. The next section shows that this change suffices to
avoid any loss in social cost.

\section{Approximating the Optimal TEF1 Factor}
\label{sec:approx}

There are now two reasons that an allocation algorithm cannot promise exact
TEF1 at low cost. A TEF1 allocation may not exist, and even when one does,
approximating the minimum social cost of a TEF1 allocation is difficult. We address both with the same objective:
approximate the optimal temporal EF1 factor, and compare social cost with that
of the cheapest allocation attaining this factor exactly.

\begin{theorem}\label{thm:fptas}
Fix the number of agents. There is an algorithm that, given an instance and
$0<\varepsilon\le1$, runs in time polynomial in the input length and
$1/\varepsilon$ and returns an allocation $\widehat\pi$ satisfying $\alpha(\widehat\pi)\le(1+\varepsilon)\alpha^*$ and $\mathrm{sc}(\widehat\pi)\le\mathrm{sc}^*$.

Moreover, given also any $\alpha\ge1$ for which an $\alpha$-TEF1 allocation
exists, an algorithm with the same running-time guarantee returns an
allocation $\pi_\alpha$ satisfying $\alpha(\pi_\alpha) \le(1+\varepsilon)\alpha$ and $\mathrm{sc}(\pi_\alpha)\le
 \min_{\pi:\,\alpha(\pi)\le\alpha}\mathrm{sc}(\pi)$.
\end{theorem}

The fairness and cost guarantees apply to the same allocation $\widehat\pi$.
In particular, when $\alpha^*=1$, the algorithm obtains
$(1+\varepsilon)$-TEF1 without exceeding the social cost of the cheapest TEF1
allocation, even though it does not decide whether a TEF1 allocation exists.
This is compatible with Theorem~\ref{thm:cost-hardness} because the output is allowed a small
fairness relaxation; setting $\varepsilon$ to zero is not part of the
algorithmic guarantee.

When $\alpha^*>1$, the theorem does not promise
$(1+\varepsilon)$-TEF1, but only a factor close to the instance's own
optimum, which can be arbitrarily large by
Theorem~\ref{thm:unbounded}. The target-factor statement is also useful
on its own: for any attainable target $\alpha$, relaxing it to
$(1+\varepsilon)\alpha$ incurs no social cost beyond that of the cheapest
$\alpha$-TEF1 allocation. The algorithm need not minimize social cost among
all $(1+\varepsilon)\alpha$-TEF1 allocations.

\subsection{Approximating the TEF1 Factor}

The algorithm processes chores in their prescribed order and retains selected
partial allocations. An exact dynamic program would need to distinguish
many different bundle costs. We instead group partial allocations whose costs
fall in the same geometric intervals and keep one allocation per group, as is
common in approximation schemes based on dynamic
programming~\citep{IbarraKim1975,HorowitzSahni1976,Woeginger2000}. The
grouping has to reflect that TEF1 compares residual costs, not the total
costs of the agents' own bundles.

Approximating the cost of an agent's own bundle and the cost of her largest
own chore separately would be insufficient. For example, if her own bundle
costs $1001$ and her largest own chore costs $1000$, her residual is $1$, but
a relative error of only $1\%$ in either number amounts to $10$.
We therefore represent each partial allocation by each agent's residual and
her costs for the other bundles, which we call its coordinates, and keep each
agent's largest own-chore cost exact. By \eqref{eq:residual-update}, two partial allocations
with the same largest own-chore costs receive exactly the same nonnegative
coordinate increments if the next chore has the same recipient. A multiplicative
comparison between their coordinates is preserved under such increments.
This lets us control the effect of replacing one partial allocation by
another in the same group at each round.

A second issue is specific to the temporal setting: a partial allocation's
current costs do not reveal how unfair an earlier prefix may have been. Each retained
allocation therefore also stores its temporal EF1 factor over all rounds so
far, computed exactly rather than estimated from grouped costs. Zero residuals
and zero comparison costs are likewise kept separate from positive values, so
that replacing an allocation by another in the same group never turns a
finite ratio into an infinite one (i.e., a positive residual compared with a
zero-cost bundle).

\subsection{Preserving Social Cost}

We use the same grouping in two dynamic programs, but choose the retained
allocation differently. The first keeps, in each group, the allocation with
the smallest temporal EF1 factor attained so far, and thereby obtains a
factor close to $\alpha^*$. The second takes a target factor, discards partial allocations
whose factor exceeds a threshold that starts at the target and increases
slightly with each round, and keeps the cheapest remaining allocation in each
group.

For the second program, fix any allocation that meets the target. We show
that, at every round, some retained partial allocation has costs close to
those of this allocation's prefix and social cost no larger, so it can be
extended by the same future recipients without exceeding the threshold.
Extending both allocations by the same recipient adds the same social cost,
and keeping the cheapest allocation in a group can then only improve the
cost comparison. This is
why fairness incurs a small multiplicative loss while social cost incurs
none. Such bicriteria guarantees are familiar, e.g.,
for generalized assignment~\citep{ShmoysTardos1993} and scheduling with
costs~\citep{AngelBampisKononov2001}; here, the relaxed criterion is fairness
at every prefix. The threshold applies to the exact factor over all rounds so
far, not only to the current partial allocation.

Finally, we use the factor achieved by the first program as the target for
the second. Every allocation attaining $\alpha^*$ meets that target, so
the output of the second program has social cost at most
$\mathrm{sc}^*$. Dividing the desired error between the two
programs yields Theorem~\ref{thm:fptas}.
Appendix~\ref{app:fptas} gives the algorithms, the full inductive arguments,
and the running-time analysis. The degree of the polynomial depends on the
number of agents, so the result does not give polynomial running time when
this number is part of the input. The algorithms also use complete
knowledge of the instance, as permitted by our informed model.

\section{Conclusion}
\label{sec:conclusion}

In this paper, we have studied temporal fair division of chores in the
informed model and resolved the existence and complexity questions left open
by \citet{ElkindEtAl2025}. Our results show that temporal
fairness for chores can be far more demanding than static EF1. For every
$n\ge3$, the optimal temporal EF1 factor can be arbitrarily large with a
fixed number of chores, and nine suffice for $n=3$. At the other extreme,
when all costs are nearly equal and a temporal EF1 factor close to $1$ is
immediate, deciding exact TEF1 existence remains NP-complete for every fixed
$n\ge3$. These results give two distinct
reasons for approximating the optimal temporal EF1 factor: some instances require
substantial relaxation, while others are difficult only at exact fairness.

Social cost gives a complementary reason. Keeping temporal fairness exact
can impose an unbounded cost increase even with two agents, unlike static
EF1, and, unless P$=$NP, minimum-cost TEF1
cannot be approximated in polynomial time within any polynomial factor, even
when a TEF1 allocation is supplied. For a fixed number of agents, however, our FPTAS
returns an allocation whose temporal EF1 factor is at most $1+\varepsilon$
times the optimum and whose social cost is at most $\mathrm{sc}^*$.

Several questions remain. With three agents, nine chores suffice both for
TEF1 nonexistence and for unbounded optimal factors
(Theorem~\ref{thm:unbounded}); what is the minimum number needed in each
case? For maximin share allocations, \citet{FeigeSapirTauber2021}
answered the analogous question for three agents and goods, as did
\citet{Teh2026Increasing} for the compatibility of EF1 and Pareto optimality
with two agents and strictly increasing valuations.
Beyond the preference classes for which TEF1 allocations are known to
exist~\citep{ElkindEtAl2025,ChoiLiTeh2026},
which restrictions on preferences guarantee a bounded optimal factor? On the
algorithmic side, can the dependence on the number of agents be improved, or
can the optimal factor be approximated efficiently when that number is part of
the input?
\subsection*{Acknowledgments}
The author would like to thank Sofia Simola for contributing to ideas for the nonexistence proof in Oct 2025. GPT-6 was used to revise the exposition and polish the proofs. The author is responsible for the final content of the paper.

\bibliographystyle{plainnat}
\bibliography{arxiv}
\clearpage
\appendix
\section{Omitted Proofs from Section~\ref{sec:nonexistence}}
\label{app:nonexistence}

We give complete proofs of the nonexistence claims in
Section~\ref{sec:nonexistence}. The family in Table~\ref{tab:powers} rules out
any prescribed multiplicative factor. A second instance, in
Table~\ref{tab:nine-chores}, establishes nonexistence with small integer costs
and provides the initial nine chores for our reductions. The two instances
have the same pattern of possible recipient sequences under the comparisons
used below. We first derive those comparisons, then provide all numerical
entries in Table~\ref{tab:complete-prefixes}.

For the calculations, a string such as $dpq$ specifies recipients in arrival
order: its first symbol is the recipient of $c_1$, its second the recipient of
$c_2$, and so on. In both instances, all costs are positive. Thus, in any
allocation with finite temporal EF1 factor, the first three chores must have
distinct recipients. Otherwise an agent has two positive-cost chores while
another agent has an empty bundle; removing one own chore leaves positive cost
compared with zero. There are therefore exactly six possible initial strings.

\subsection{The Three-Agent Case of Theorem~\ref{thm:unbounded}}
\label{app:unbounded-proof}

\begin{proof}
Fix an integer $h\ge2$ and use Table~\ref{tab:powers}. We prove that no
$(h-1)$-TEF1 allocation exists. Let an agent own at least two chores at some
prefix, and suppose that her second-largest own cost is $h^k$. Her residual
is at least $h^k$. If some other bundle contains only chores that she values
strictly below $h^k$, then its cost is at most
$1+h+\cdots+h^{k-1}=(h^k-1)/(h-1)<h^k/(h-1)$,
since each power smaller than $h^k$ occurs at most once in her row (the only
repeated power, $h^7$, is her largest cost). For $k=0$, this sum is empty and equals
zero, and the same conclusion follows. Consequently, her inequality in the
definition of $(h-1)$-TEF1 is violated at that prefix.

It follows that every $(h-1)$-TEF1 allocation must satisfy the following
necessary condition at every prefix: an agent's second-largest own cost is no
greater than her largest cost for a chore in each other bundle. Since every
cost is a power of $h$, we compare these costs through their exponents. Once the first
three chores have distinct recipients, all of these exponents are defined
whenever we consider giving an agent a second chore. Only the recipient needs
to be checked when a chore arrives. Every other agent's own exponents stay
fixed, while the largest exponent in a comparison bundle can only increase.

The three columns headed ``Exponent comparison'' in
Table~\ref{tab:complete-prefixes} apply this necessary condition. The entry for a recipient $i$ is the minimum,
over $j\ne i$, of the largest exponent in $A_j$ according to $i$, minus the
second-largest exponent among $i$'s own chores after she receives the next
chore. A negative entry excludes that recipient. Starting with the six strings of
length three and retaining exactly the nonnegative choices, the table includes
every possible continuation under the necessary condition. After eight
rounds, only $dpqpqdpp$, $dpqqpqdq$, and $dpqqdpqd$ remain. For each of these strings, every entry for the ninth chore is negative.
Hence there is no allocation satisfying the condition through all nine rounds,
and therefore no $(h-1)$-TEF1 allocation.

For an arbitrary prescribed $\alpha\ge1$, choose an integer $h\ge\alpha+1$.
An $\alpha$-TEF1 allocation would also be $(h-1)$-TEF1, which we have ruled
out. All rows have the same multiset of costs and the same positive total.
Dividing the entire matrix by this total normalizes all rows and preserves
every fairness ratio. This proves the theorem for $n=3$ and $m=9$.
\end{proof}

The geometric comparison explains why the argument uses the second-largest
own chore. EF1 removes one chore, so making only the largest own chore costly
would not be enough. In contrast, once a second own chore exceeds every chore
in a comparison bundle, its cost cannot be removed along with the largest
one. The finite calculation establishes that the given arrival order forces
such a comparison somewhere, regardless of the allocation.

\subsection{Nonexistence with Costs at Most Ten}
\label{app:small-instance}

We next prove that the instance in Table~\ref{tab:nine-chores} admits no
TEF1 allocation. For an EF1 prefix
$A=A^{t-1}$ and a proposed recipient $i$ of $c_t$, the next prefix is EF1
exactly when
\begin{equation}\label{eq:extension-test}
 \min_{j\ne i}d_i(A_j)-R_i(A)-\min\{M_i(A),d_i(c_t)\}\ \ge\ 0.
\end{equation}
Indeed, by \eqref{eq:residual-update}, the left-hand side is $i$'s least
comparison cost minus her new residual, and, as noted in
Section~\ref{sec:model}, no other agent can newly violate EF1. Thus it
suffices to compute the expression in \eqref{eq:extension-test} for each
possible recipient.

The last three columns of Table~\ref{tab:complete-prefixes} give exactly these
values. The six length-three strings produce six strings of length four,
then six of length five, five of length six, five of length seven, and three
of length eight. All entries are negative at the ninth chore, so no TEF1
allocation exists. This argument checks all possible partial allocations,
not just those produced by one particular rule.

For example, after the first three recipients $pdq$, the own costs are
$2,3,1$. If $c_4$ goes to $p$, her new residual is $2$, whereas she evaluates
$d$'s bundle at $1$, giving the entry $-1$. If it goes to $d$, her residual is
$3$, whereas she evaluates $q$'s bundle at $1$, giving $-2$. If it goes to
$q$, her residual is $1$ and her least costly comparison bundle costs $2$,
giving $1$. Thus only $q$ may receive $c_4$. Every other row is obtained in
exactly the same way.

The cost sets of agents $p,d,q$ are respectively
$\{1,2,4,8\}$, $\{1,2,3,5\}$, and $\{1,2,5,10\}$. Each has four distinct
costs, and the last two columns are identical. The example therefore has
eight chore types, where a type is a complete vector of costs across agents.

\subsection{Complete Prefix Calculation}
\label{app:prefix-calculation}

Table~\ref{tab:complete-prefixes} can be read one group at a time. Start with
all six permutations of $p,d,q$. In each row, append precisely those agent
labels whose entries are nonnegative. These extensions, and no others, form
the next group. Thus every partial allocation that could meet the relevant
fairness requirement is considered again when the next chore arrives.

For a direct check near the end, consider the string $dpqqpqdq$. Its bundles
before the ninth chore are $A_p=\{c_2,c_5\}$, $A_d=\{c_1,c_7\}$, and
$A_q=\{c_3,c_4,c_6,c_8\}$.
In the instance of Table~\ref{tab:powers}, assigning $c_9$ to $p$ leaves a second-largest
own cost of $h^5$, while her largest cost in $A_d$ is only $h^4$.
Assigning it to $d$ gives the same comparison against $A_p$.
Assigning it to $q$ leaves a second-largest own cost of $h^7$, while her
largest cost in $A_p$ is $h^5$. The corresponding entries are therefore
$-1,-1,-2$. All three choices are excluded without estimating any residual
sum. The other two rows of the last group exclude all recipients as well.

The nonnegative choices are the same in the two sets of columns. They have
different meanings, however: for the instance of Table~\ref{tab:powers},
they are only necessary for $(h-1)$-TEF1, while for the instance of
Table~\ref{tab:nine-chores}, they are exactly the choices that keep the next
prefix EF1. Allowing more partial allocations than fairness permits
does not affect the nonexistence proof when even this larger collection has
no continuation.

\begin{table}[t]
\caption{All continuations for the two nine-chore instances. The first three
numerical columns give the exponent comparison in the proof of
Theorem~\ref{thm:unbounded}. The last three give the expression in
\eqref{eq:extension-test} for Table~\ref{tab:nine-chores}. In either set of
columns, a recipient may be retained only when its entry is nonnegative.
For the fixed numerical instance this condition is also sufficient. Each
subsequent group includes exactly the strings obtained from the preceding
group.}
\label{tab:complete-prefixes}
\centering
\begin{tabular}{cc|rrr|rrr}
\toprule
 & & \multicolumn{3}{c|}{Exponent comparison} & \multicolumn{3}{c}{EF1 inequality}\\
Next chore & Current recipients & $p$ & $d$ & $q$ & $p$ & $d$ & $q$\\
\midrule
$c_4$ & $pdq$ & $-2$ & $-2$ & $1$ & $-1$ & $-2$ & $1$ \\
$c_4$ & $pqd$ & $-2$ & $1$ & $-1$ & $-1$ & $1$ & $-1$ \\
$c_4$ & $dpq$ & $1$ & $-1$ & $1$ & $1$ & $-1$ & $1$ \\
$c_4$ & $dqp$ & $-1$ & $-1$ & $-1$ & $-1$ & $-1$ & $-1$ \\
$c_4$ & $qpd$ & $1$ & $1$ & $-2$ & $1$ & $1$ & $-1$ \\
$c_4$ & $qdp$ & $-1$ & $-2$ & $-2$ & $-1$ & $-2$ & $-1$ \\
\midrule
$c_5$ & $pdqq$ & $-2$ & $-1$ & $-1$ & $-1$ & $-1$ & $-1$ \\
$c_5$ & $pqdd$ & $-2$ & $-2$ & $1$ & $-1$ & $-2$ & $1$ \\
$c_5$ & $dpqp$ & $-2$ & $-1$ & $2$ & $-1$ & $-1$ & $3$ \\
$c_5$ & $dpqq$ & $2$ & $1$ & $-1$ & $1$ & $1$ & $-1$ \\
$c_5$ & $qpdp$ & $-2$ & $1$ & $-3$ & $-1$ & $1$ & $-4$ \\
$c_5$ & $qpdd$ & $2$ & $-2$ & $-2$ & $1$ & $-2$ & $-3$ \\
\midrule
$c_6$ & $pqddq$ & $1$ & $-2$ & $-2$ & $2$ & $-2$ & $-4$ \\
$c_6$ & $dpqpq$ & $-1$ & $2$ & $-2$ & $-1$ & $2$ & $-2$ \\
$c_6$ & $dpqqp$ & $-3$ & $2$ & $1$ & $-7$ & $2$ & $2$ \\
$c_6$ & $dpqqd$ & $3$ & $-2$ & $-1$ & $3$ & $-2$ & $-1$ \\
$c_6$ & $qpdpd$ & $-1$ & $-3$ & $-1$ & $-1$ & $-2$ & $-1$ \\
$c_6$ & $qpddp$ & $-3$ & $-2$ & $-1$ & $-7$ & $-2$ & $-2$ \\
\midrule
$c_7$ & $pqddqp$ & $-1$ & $1$ & $-3$ & $-2$ & $2$ & $-9$ \\
$c_7$ & $dpqpqd$ & $2$ & $-2$ & $-3$ & $7$ & $-3$ & $-7$ \\
$c_7$ & $dpqqpd$ & $-1$ & $-2$ & $2$ & $-1$ & $-3$ & $7$ \\
$c_7$ & $dpqqpq$ & $-2$ & $3$ & $-1$ & $-3$ & $4$ & $-3$ \\
$c_7$ & $dpqqdp$ & $-1$ & $-1$ & $2$ & $-1$ & $-1$ & $4$ \\
\midrule
$c_8$ & $pqddqpd$ & $-2$ & $-1$ & $-1$ & $-2$ & $-3$ & $-2$ \\
$c_8$ & $dpqpqdp$ & $1$ & $-2$ & $-1$ & $3$ & $-3$ & $-1$ \\
$c_8$ & $dpqqpdq$ & $-1$ & $-2$ & $-2$ & $-1$ & $-1$ & $-3$ \\
$c_8$ & $dpqqpqd$ & $-1$ & $-1$ & $1$ & $-3$ & $-1$ & $4$ \\
$c_8$ & $dpqqdpq$ & $-2$ & $1$ & $-2$ & $-1$ & $3$ & $-6$ \\
\midrule
$c_9$ & $dpqpqdpp$ & $-2$ & $-2$ & $-1$ & $-5$ & $-3$ & $-1$ \\
$c_9$ & $dpqqpqdq$ & $-1$ & $-1$ & $-2$ & $-3$ & $-1$ & $-6$ \\
$c_9$ & $dpqqdpqd$ & $-2$ & $-2$ & $-2$ & $-1$ & $-2$ & $-6$ \\
\bottomrule
\end{tabular}
\end{table}
\FloatBarrier

\subsection{A Unique Initial Allocation for the Reductions}
\label{app:unique-prefix}

A slight modification of the small instance has a unique TEF1 allocation, so
the reductions can fix the assignment of their first nine chores without
specifying it as part of the input. We state the modification with a scaling
factor $B$ and a nonnegative offset $s$, because the two reductions use
different values of $B$ and $s$.

\begin{lemma}\label{lem:unique-prefix}
Let $B$ be a positive integer and $s\ge0$. Multiply the first eight columns
of Table~\ref{tab:nine-chores} by $B$, and give the ninth chore cost vector $(8B+s,\ 4B,\ 10B+s)$.
The resulting instance has a unique TEF1 allocation, with recipient string
$dpqqpqdqd$. Its bundle-cost matrix is
\begin{equation}\label{eq:prefix-matrix}
\begin{array}{c|ccc}
 & A_p & A_d & A_q\\\hline
p & 9B & 14B+s & 20B\\
d & 6B & 11B & 14B\\
q & 12B & 25B+s & 18B
\end{array}
\end{equation}
and its vectors of maximum own costs and residuals are respectively
$(8B,5B,10B)$ and $(B,6B,8B)$.
\end{lemma}

\begin{proof}
Common scaling preserves all EF1 inequalities, so the TEF1 recipient strings
for the first eight chores are exactly the three strings of length eight in
Table~\ref{tab:complete-prefixes}.
In each, the maximum own costs are $(8B,5B,10B)$. Before the ninth chore,
the differences between least comparison cost and own residual, in the
order $(p,d,q)$, are $(3B,2B,9B)$, $(5B,4B,4B)$, and $(7B,3B,4B)$ for the
strings $dpqpqdpp$, $dpqqpqdq$, and $dpqqdpqd$, respectively.
Giving the ninth chore to $p$ increases her residual by $8B$, regardless
of $s$, because her maximum before the assignment is $8B$. Giving it to
$q$ similarly increases the residual by $10B$. Neither increase is
permitted in any of the three allocations. Giving it to $d$ increases
her residual by $4B$, which is permitted only in the middle allocation.
The unique resulting string is therefore $dpqqpqdqd$.

Its bundles are $A_p=\{c_2,c_5\}$, $A_d=\{c_1,c_7,c_9\}$, and
$A_q=\{c_3,c_4,c_6,c_8\}$.
Adding the corresponding entries gives \eqref{eq:prefix-matrix}. The
largest costs in the own bundles are $8B,5B,10B$, and subtracting them
from the diagonal entries gives the stated residuals.
\end{proof}

\subsection{More Than Three Agents}
\label{app:general-nonexistence}

We extend the geometric construction to any number of at least three
agents. The following lemma explains how to make $n-3$ agents receive only
one chore each, without specifying their identities in advance.

\begin{lemma}\label{lem:initial-recipients}
Let $n\ge3$ and $\alpha\ge1$. Consider a nonempty list of chores with strictly positive
costs, and assign a nonnegative integer $e_i(c)$ to every agent $i$ and
chore $c$. Suppose that, for each $i,c$, at most two preceding chores $c'$
satisfy $e_i(c')\ge e_i(c)$, and
\begin{equation}\label{eq:lower-exponent-sum}
 d_i(c)>
 \alpha\sum_{\substack{c'\text{ precedes }c\\ e_i(c')<e_i(c)}}d_i(c').
\end{equation}
Prepend $n-3$ chores, each of cost $F_i>\max_c d_i(c)$ to agent $i$.
In every $\alpha$-TEF1 allocation of the resulting instance, these initial
chores have distinct recipients, and none of their recipients receives a
later chore.
\end{lemma}

\begin{proof}
Positivity forces distinct recipients for the first $n-3$ chores: a repeated
recipient would have positive residual while another bundle is empty. Let
$T$ be the three agents who receive none of these chores. Suppose that $c$
is the first later chore assigned to an agent $i\notin T$. Before this
assignment, her bundle contains only an initial chore of cost $F_i>d_i(c)$,
so her new residual is $d_i(c)$. At most two preceding later chores $c'$ satisfy $e_i(c')\ge e_i(c)$. Hence some agent $j\in T$
owns none of them. Her bundle contains no initial chore either, so its cost
to $i$ is at most the sum in \eqref{eq:lower-exponent-sum}. This contradicts
$i$'s $\alpha$-EF1 inequality toward $j$.
\end{proof}

\begin{proof}[Completion of the proof of Theorem~\ref{thm:unbounded}]
Let $r_\tau(t)$ be the exponent in row $\tau\in\{p,d,q\}$ and column $t$ of
Table~\ref{tab:powers}. In every row, each column has at most two earlier
columns whose exponents are at least its own.

For each $n\ge3$, choose maps
$f_1,\ldots,f_{q_n}\colon[n]\to\{p,d,q\}$ such that every three-agent set
$T$ satisfies $f_b(T)=\{p,d,q\}$ for some $b$. We can choose $q_3=1$ and
$q_n=O(\log n)$. Indeed, if the labels of all agents are chosen independently
and uniformly, the three labels on a given triple are distinct with
probability $6/27=2/9$. For $\ell$ independently chosen maps, the probability
that some triple never receives three distinct labels is at most
$\binom n3(7/9)^\ell$, which is less than $1$ when
$\ell>\log_{9/7}\binom n3$.
Thus such a family exists with
$q_n\le1+\lceil\log_{9/7}\binom n3\rceil$, independently of $\alpha$.

Set $m_n=n-3+9q_n$ and choose an integer $h$ with $h-1>2\alpha$.
Begin with $n-3$ chores of cost $h^{8q_n}$ to every agent. For each
$b=0,\ldots,q_n-1$, append a block of nine chores $c_{b,1},\dots,c_{b,9}$,
where $c_{b,t}$ costs $h^{e_i(c_{b,t})}$ to agent $i$, with exponent
$e_i(c_{b,t})=8b+r_{f_{b+1}(i)}(t)$.
The exponents in successive blocks lie in disjoint increasing intervals.
Each exponent occurs at most twice in a row among these blocks. Therefore,
for every nonnegative integer $e$,
\begin{equation}\label{eq:general-geometric}
 \sum_{\substack{c\text{ in a block}\\ e_i(c)<e}}d_i(c)
 \le 2\sum_{\ell=0}^{e-1}h^\ell
 <\frac{2h^e}{h-1}<\frac{h^e}{\alpha},
\end{equation}
where the sum is zero when $e=0$.
For each arriving block chore, at most two earlier block chores have an
exponent at least its own. The initial cost $h^{8q_n}$ exceeds every block
cost. Lemma~\ref{lem:initial-recipients} therefore implies that only the
three agents who receive no initial chore can receive block chores.

Suppose an $\alpha$-TEF1 allocation exists, and let $T$ be these three
agents. Choose a block whose labels on $T$ are $p,d,q$. Its first three
chores must have distinct recipients. Otherwise an agent owns two chores
from this block while another agent in $T$ owns none; after removing one own
chore, a remaining block chore costs more than $\alpha$ times the entire
comparison bundle, by \eqref{eq:general-geometric}.

Now consider any later prefix within this block. If an agent's
second-largest own exponent among the block chores exceeds her largest
exponent in another agent's bundle restricted to the block, let $e$ be
that second-largest exponent. Her residual is at least $h^e$. All chores
in the comparison bundle, including those from earlier blocks, have smaller
exponents, so the bundle costs less than $h^e/\alpha$ by
\eqref{eq:general-geometric}. Thus the allocation restricted to this block
must satisfy the necessary exponent comparisons of
Appendix~\ref{app:unbounded-proof}. Table~\ref{tab:complete-prefixes} shows
that no such allocation exists through its ninth chore, a contradiction.

Each block has the same multiset of individual costs in every row, as do
the initial chores. All row totals are therefore equal. Appending identical
positive-cost columns gives every $m\ge m_n$ without restoring an
$\alpha$-TEF1 allocation or changing equality of the row multisets.
Dividing by the common row total normalizes the costs and preserves every
fairness ratio. Finally, $m_n=n+O(\log n)$ and $m_3=9$, as required.
\end{proof}

\section{Omitted Proofs from Section~\ref{sec:uniform}}
\label{app:uniform-proofs}

\subsection{NP-Completeness without the Cost Restriction}
\label{app:hardness}

We first prove that \textsc{TEF1-Chores} is NP-complete for three agents,
without the restriction on costs in Theorem~\ref{thm:near-uniform}. Unlike in
the problem of completing a given partial allocation, the first nine
recipients are not specified in the input; instead,
Lemma~\ref{lem:unique-prefix} forces them. After that initial allocation, the reduction makes two agents divide
chores that have the same cost to both of them.

\begin{proof}[Proof of three-agent NP-completeness]
Membership in NP follows by checking the residual inequalities at all
prefixes of a proposed recipient sequence. For hardness, reduce from
\partprob{}~\citep{GareyJohnson1979}. Let $a_1,\dots,a_r$ be positive
integers, write $U=\sum_{j=1}^r a_j$, and set $B=2^{r+3}U$.
Use the nine-chore instance of Lemma~\ref{lem:unique-prefix} with $s=0$.
Append an input chore $x_j$ for each number $a_j$, with cost vector
$(2^{r+4}a_j,\,2^{j-1},\,2^{r+4}a_j)$ for $j=1,\dots,r$.
Finally append four chores with cost vectors $(6B,2^r,6B)$,
$(8B,2^{r+1},8B)$, $(6B,2^{r+2},6B)$, and $(8B,2^{r+3},8B)$, in this order.
The construction has $r+13$ chores, uses positive integer costs, and takes
polynomial time. We refer to the $r+4$ chores after the first nine as the
\emph{later chores}; each of them has the same cost to $p$ and $q$.

\paragraph{The later chores cannot go to agent $d$.}
In arrival order, the later chores cost $1,2,4,\dots,2^{r+3}$ to $d$.
Suppose inductively that $d$ receives none of the
first $\ell-1$ later chores. Her residual is still $6B$ and her maximum
own cost is still $5B$. The preceding later chores have total cost
$2^{\ell-1}-1$ to her, so \eqref{eq:prefix-matrix} implies that she values
$p$'s bundle at most $6B+2^{\ell-1}-1$. On the other hand, the $\ell$th
later chore costs $2^{\ell-1}\le2^{r+3}\le B<5B$ to $d$. By
\eqref{eq:residual-update}, giving it to her would increase her residual to
$6B+2^{\ell-1}$, which exceeds her cost for $p$'s bundle. Hence $d$ cannot
receive the chore, completing the induction.

\paragraph{Any TEF1 allocation yields a partition.}
All later chores must therefore go to $p$ or $q$. At any prefix after the
first nine chores, let $T_p$ and $T_q$ be the total costs of the later chores
given to $p$ and $q$, respectively. The maximum own costs of $p$ and $q$ remain
$8B,10B$: an input chore costs at most $2B$, and a final chore costs at
most $8B$. Comparing each with the unchanged bundle of $d$ gives
$B+T_p\le14B$ and $8B+T_q\le25B$, so $T_p\le13B$ and $T_q\le17B$. The input chores have total cost $2B$
and the final four have total cost $28B$. Thus, after all later chores,
$T_p+T_q=30B$, which is the sum of the two upper bounds. Both bounds must
therefore hold with equality, i.e., $T_p=13B$ and $T_q=17B$.

Let $X$ be the total input-chore cost assigned to $p$. Since
$0\le X\le2B$, the final chores assigned to $p$ have total cost
$13B-X\in[11B,13B]$. Each subset sum of the costs
$6B,8B,6B,8B$ is an even multiple of $B$. The only possible value in that
interval is $12B$, so $X=B$. Since each input chore $x_j$ costs $2^{r+4}a_j$
to $p$, this means $2^{r+4}\sum_{j:\,x_j\in A_p}a_j=B=2^{r+3}U$, so the
numbers $a_j$ with $x_j\in A_p$ sum to $U/2$.

\paragraph{Any partition yields a TEF1 allocation.}
Conversely, suppose that such a subset exists. Give its input chores to
$p$ and all other input chores to $q$. Each then receives input cost
$B$. Give the four final chores to $p,q,p,q$, respectively. We check all
prefixes, since EF1 of the final allocation alone would not suffice.

Agent $d$ receives no later chore, so her own residual does not change,
while her comparison costs only increase. For $p$ and $q$, the four
nontrivial EF1 inequalities after any later prefix are
\begin{equation}\label{eq:four-constraints}
 T_p\le13B,\quad T_q\le17B,\quad
 T_p-T_q\le19B,\quad T_q-T_p\le4B.
\end{equation}
The first two are the comparisons with $d$. The third follows from
$B+T_p\le20B+T_q$, and the fourth from
$8B+T_q\le12B+T_p$, using \eqref{eq:prefix-matrix}.
While the input chores arrive, both $T_p$ and $T_q$ stay in $[0,B]$, so all
four inequalities hold. After each of the four final chores, the pair
$(T_p,T_q)$ is, in turn, $(7B,B)$, $(7B,9B)$, $(13B,9B)$, and $(13B,17B)$.
Substituting each pair into \eqref{eq:four-constraints} verifies all
remaining prefixes. The initial nine chores are TEF1 by
Lemma~\ref{lem:unique-prefix}. Thus the full allocation is TEF1.

We have proved that the constructed instance admits a TEF1 allocation if and only if
the \partprob{} instance has a solution. Together with NP membership,
this proves that \textsc{TEF1-Chores} is NP-complete for three agents.
\end{proof}

\subsection{More Than Three Agents}
\label{app:add-agents}

We next extend the existence problem to any fixed number of agents. The
added chores must do more than leave an arbitrary triple to receive the
original chores: they must leave the original three agents. We also arrange
that these agents are envy-free toward one another before the original
chores arrive.

\begin{lemma}\label{lem:add-agents}
For every fixed $n\ge3$, a three-agent instance with strictly positive
integer costs can be transformed in polynomial time into an $n$-agent
instance with strictly positive integer costs that admits a TEF1 allocation
if and only if the original instance does.
\end{lemma}

\begin{proof}
For $n=3$, use the original instance. Fix $n\ge4$, let $S=\{1,2,3\}$ be the
original agents, and write their chore list as $(c_1,\ldots,c_k)$. The
constructed costs are denoted by $\widetilde d_i$.

\paragraph{The added chores.}
There are $Q=\binom n3-1$ triples $T\ne S$. For each such triple, choose
labels in $\{p,d,q\}$ for all agents so that $T$ receives three distinct
labels and $S$ receives two labels $p$ and one label $d$. This is always
possible. When $|T\cap S|=2$, give those two agents labels $p,d$ and the
agent in $T\setminus S$ label $q$. When $|T\cap S|=1$, give the common
agent label $p$ and the other two members of $T$ labels $d,q$. When the
intersection is empty, label the members of $T$ arbitrarily with $p,d,q$.
In each case, complete the labels on $S$ to $p,p,d$, and give any remaining
agents label $p$.

Use one ten-chore block for each triple, in any fixed order. Put $G=10Q$
and $b=2G+1$. In block $a$, where $a=0,\ldots,Q-1$, the
first seven chores cost $b^{8a+r_\tau(t)}$ to an agent with label $\tau$,
using the exponent rows $r_\tau$ from
Appendix~\ref{app:general-nonexistence}. Write $P_1,P_2$ for the two agents
of $S$ with label $p$, and $D$ for the one with label $d$; these names apply
only within the current block. Put $H_a=b^{8a+7}$. The last three chores
have costs
\begin{equation}\label{eq:three-final-costs}
\begin{array}{c|ccc}
 &8&9&10\\ \hline
 P_1&H_a&2H_a&2H_a\\
 P_2&2H_a&H_a&2H_a\\
 D&2H_a&2H_a&H_a
\end{array}
\end{equation}
to the agents of $S$, and cost $H_a$ each to every agent outside $S$.
Use exponent $8a+7$ for each of these last three chores. Every block cost
is thus either $b^{e_i(c)}$ or $2b^{e_i(c)}$, where $e_i(c)$ is its
specified exponent.

Let $\mathcal B$ be the set of all $G$ block chores, and set $P=\max_i\widetilde d_i(\mathcal B)$, $L=P+1$.
After the blocks, append chores $x_1,\ldots,x_k$ corresponding to the
original chores, with costs $\widetilde d_i(x_t)=Ld_i(c_t)$ for $i\in S$ and
$\widetilde d_i(x_t)=L2^{t-1}$ for $i\notin S$.
Finally, prepend $n-3$ chores, each of cost
$F=1+\max_i\bigl(\widetilde d_i(\mathcal B)+\sum_{t=1}^k\widetilde d_i(x_t)\bigr)$
to every agent. All other costs are fixed before $F$ is chosen. The complete
arrival order consists of these $n-3$ chores, the $G$ block chores, and
$x_1,\ldots,x_k$.

\paragraph{Only the original agents can receive the original chores.}
Each of the $G$ block chores $c$ costs at most $2b^{e_i(c)}$ to agent $i$, and
$b=2G+1$. Hence, for every agent $i$ and every nonnegative integer $e$,
\begin{equation}\label{eq:added-agent-lower-costs}
 \sum_{c\in\mathcal B:\,e_i(c)<e}\widetilde d_i(c)\le 2Gb^{e-1}<b^e,
\end{equation}
where the sum is empty when $e=0$. In each
block, every arriving chore has at most two earlier chores with exponent at least its own: this follows from Table~\ref{tab:powers} for the
first nine chores, and the tenth has exactly two such predecessors. Earlier
blocks have smaller exponents. Lemma~\ref{lem:initial-recipients},
applied to the prefix ending after the blocks with $\alpha=1$, implies
that all block chores go to the three agents who receive no initial chore.

Let $T$ be those agents. If $T\ne S$, consider its corresponding block.
On $T$, the exponents of its first nine chores are precisely the
three exponent rows of Table~\ref{tab:powers}, plus $8a$. The inequality
\eqref{eq:added-agent-lower-costs} implies, exactly as in
Appendix~\ref{app:general-nonexistence}, that restricting a TEF1 allocation
to these nine chores must satisfy the exponent comparisons in
Table~\ref{tab:complete-prefixes}. No such allocation exists. Hence $T=S$,
and every agent outside $S$ owns exactly one initial chore after the blocks.

Suppose $x_t$ is the first appended chore assigned to an agent $i\notin S$.
Her largest own cost remains $F$, so her new residual is $L2^{t-1}$.
All preceding block and appended chores together cost her at most
$P+L\sum_{s=1}^{t-1}2^{s-1}=L2^{t-1}-1$.
Every bundle of an agent in $S$ is a subset of these chores, contradicting
EF1. Thus every $x_t$ must go to an agent in $S$.

\paragraph{A TEF1 allocation of the added prefix.}
Give the first $n-3$ chores to distinct agents outside $S$. Within each
block, assign the ten chores in the order
$D,P_1,P_2,P_1,P_2,D,P_1,P_1,P_2,D$.
We prove inductively that the allocation is TEF1 and that, after every
complete block, each agent in $S$ values her own bundle less than the
bundle of either other agent in $S$.

The bundles of $S$ are initially empty. At the beginning of any block,
they are therefore envy-free, either by emptiness or by the induction
hypothesis. The first three block chores have distinct recipients, so
removing the newly received chore leaves the recipient's previous own
bundle, which costs no more than either previous comparison bundle. These
three assignments preserve EF1 among $S$.

For the remaining assignments, subtract the common block offset $8a$ from
the exponents. The second-largest own exponent among current-block
chores and the least of the two largest comparison exponents within the
block are as follows:
\[
\begin{array}{c|rrrrrrr}
 \text{chore}&4&5&6&7&8&9&10\\ \hline
 \text{second-largest own exponent}&0&1&1&3&4&5&6\\
 \text{least comparison maximum}&1&2&3&5&5&6&7
\end{array}
\]
Every entry in the second row is strictly smaller than the corresponding
entry in the third. After removing a largest own chore, all remaining own
chores, including those from earlier blocks, have exponents less
than the largest exponent in either comparison bundle. Inequality
\eqref{eq:added-agent-lower-costs} therefore gives the recipient's two EF1
inequalities. Nonrecipients' inequalities remain valid.

At the end of a block, each member of $S$ owns exactly one of its last
three chores, at cost $H_a$ to herself. All her other own chores together
cost less than $H_a$ by \eqref{eq:added-agent-lower-costs}. Each other bundle
in $S$ contains a chore costing $2H_a$ to her, by
\eqref{eq:three-final-costs}. Hence her own bundle costs less than either
comparison bundle, completing the induction. Each agent outside $S$
always has a singleton and residual zero. Comparisons from $S$ to these
agents hold throughout because $F$ exceeds every agent's total cost for all
block and appended chores.

\paragraph{A TEF1 allocation of the constructed instance yields one of the original instance.}
Take any TEF1 allocation of the constructed instance. Let $B_1,B_2,B_3$ be
the bundles of $S$ after the blocks, and put
$b_{ij}=\widetilde d_i(B_j)$ for $i,j\in S$. Every $b_{ij}$ lies in $[0,P]$.
Keep the recipient of $x_t$ as the recipient of $c_t$ in the original
instance. We have shown that all these recipients belong to $S$.

Consider a round in which $x_t$ goes to $i\in S$, and let $A$ be the
corresponding prefix of the original allocation. Since $L>P$ and original
costs are positive integers, $i$'s largest own appended chore costs more
than every own block chore. Her constructed EF1 inequality toward
$j\in S\setminus\{i\}$ is therefore $b_{ii}+LR_i(A)\le b_{ij}+Ld_i(A_j)$.
It follows that $L\bigl(R_i(A)-d_i(A_j)\bigr)\le b_{ij}-b_{ii}\le P<L$.
The quantity in parentheses is an integer, so it is nonpositive. Thus the
recipient satisfies original EF1. Nonrecipients' original own bundles are
unchanged and their comparison costs do not decrease. Induction from the
empty allocation proves original TEF1 at every prefix.

\paragraph{A TEF1 allocation of the original instance extends to the constructed instance.}
Use the allocation of the added prefix constructed above.
Its final bundles satisfy $b_{ii}<b_{ij}$ for all distinct $i,j\in S$.
Assign each $x_t$ to the original recipient of $c_t$. Adding
$b_{ii}\le b_{ij}$ to the scaled original inequality
$L R_i(A)\le Ld_i(A_j)$ proves the constructed recipient's EF1 inequality
toward each other agent in $S$. The formula for her residual is valid for
the same reason as above. Nonrecipients' inequalities cannot become
violated. All comparisons involving an agent outside $S$ hold because
that agent retains residual zero and each agent in $S$ values that singleton
at $F$, more than her total cost for all block and appended chores. Hence
the full allocation is TEF1.

The construction has $k+n-3+G$ chores. For fixed $n$, the values
$G,b,P,L$ depend only on $n$. Each cost $L2^{t-1}$ has $O(t+\log L)$ bits,
and the scaled original costs and $F$ have polynomial binary length.
All costs are positive integers, and the transformation takes polynomial
time.
\end{proof}

Together with Appendix~\ref{app:hardness}, Lemma~\ref{lem:add-agents}
establishes NP-completeness for every fixed $n\ge3$ before any restriction
on the ratio between costs is imposed.

\subsection{An Existence-Preserving Transformation}
\label{app:near-uniform}

We now transform a positive-cost instance into one whose costs are close to
a common value. Adding a common amount to each cost without changing the
arrival sequence would not suffice: EF1 comparisons also depend on the
numbers of chores in the bundles. We therefore also insert chores that cost
the same to every agent, so that all bundles have the same size just before
each original chore after the first $n$ arrives.

Take a positive-cost instance with $n$ agents and $m>n$ chores. Write
$\widehat m$ for the number of chores in the constructed instance. Let
$S=\max_i d_i(C)$, and choose an integer $K>S$. Replace each original chore $c_t$ by a chore
$\widehat c_t$ whose cost to agent $i$ is $\widehat d_i(\widehat c_t)=K+d_i(c_t)$.
An additional chore has cost $K$ to every agent. Write $u$ for such a
chore; every occurrence of $u$ below denotes a distinct chore. The new
arrival order is
\begin{equation}\label{eq:uniform-sequence}
 \widehat c_1,\dots,\widehat c_n,\quad
 (\widehat c_{n+1},u^{n-1}),\quad\dots,\quad
 (\widehat c_{m-1},u^{n-1}),\quad\widehat c_m,
\end{equation}
where $u^{n-1}$ means $n-1$ additional chores in succession. When
$m=n+1$, there are no parenthesized groups. The number of chores is $\widehat m=n+n(m-n-1)+1=n(m-n)+1$.

\paragraph{In every EF1 prefix, bundle sizes differ by at most one.}
Let $\widehat A$ be an EF1 partial allocation of the constructed instance,
and write $k_i=|\widehat A_i|$. If $k_i\ge k_j+2$, then even after removing
one own chore, agent $i$ has remaining cost at least $(k_i-1)K\ge(k_j+1)K$.
Each chore in $\widehat A_j$ costs $K$ to agent $i$, plus its original cost
if it is an original chore, and her original costs sum to at most $S$. Hence
$\widehat d_i(\widehat A_j)\le k_jK+S<(k_j+1)K$, contradicting EF1. Thus all bundle sizes differ by at most one. Whenever
the total number of assigned chores is divisible by $n$, they must be
equal. In particular, they are equal after the first $n$ chores and
after every complete parenthesized group in
\eqref{eq:uniform-sequence}.

\paragraph{A TEF1 allocation of the constructed instance yields one of the original instance.}
Suppose the constructed instance admits a TEF1 allocation. Keep the
recipients of
$\widehat c_1,\dots,\widehat c_m$ and discard the additional chores.
This defines an allocation of the original instance. The first $n$
recipients are distinct, by positivity, so all its prefixes through
round $n$ are EF1.

We proceed inductively. Immediately before $\widehat c_t$, for $t>n$,
all constructed bundle sizes are equal, say to $b$. Suppose this chore
goes to agent $i$. After the assignment, her bundle has size $b+1$ and
every other bundle has size $b$. Let $A=A^t$ be the prefix at round $t$ of
this original allocation. Every agent already received an original
chore among the first $n$ arrivals, and original costs are positive.
Consequently, $i$'s most costly own constructed chore costs
$K+M_i(A)$; an additional chore, of cost $K$, cannot cost more. Her
constructed residual and her constructed cost for the bundle of each agent
$j\ne i$ are therefore $bK+R_i(A)$ and $bK+d_i(A_j)$, respectively.
Cancelling the common term $bK$ in her constructed EF1 inequality gives
$R_i(A)\le d_i(A_j)$, exactly the recipient's original EF1 inequality.
For every nonrecipient, original EF1 is preserved because her own bundle
is unchanged and her comparison costs do not decrease. By induction, the
original allocation is TEF1.

\paragraph{A TEF1 allocation of the original instance extends to the constructed instance.}
Conversely, suppose the original instance admits a TEF1 allocation.
Give each $\widehat c_t$ to the original recipient of $c_t$. The first
$n$ recipients are distinct. In each parenthesized group, give the
$n-1$ additional chores, one each and in any order, to the $n-1$ agents
other than the recipient of the group's original chore. Bundle sizes then differ by at most one at every prefix, not
only at the end of each group.

The first $n$ constructed prefixes are EF1 because every nonempty
bundle is a singleton. At a later prefix, let $A$ be the corresponding
prefix of the original allocation and let $k_i$ be the constructed bundle
sizes.
Each agent owns an original chore. Her constructed maximum is
$K+M_i(A)$, so her residual is $(k_i-1)K+R_i(A)$.
Her cost for $j$'s bundle is $k_jK+d_i(A_j)$. Original TEF1 gives
$R_i(A)\le d_i(A_j)$, and balanced sizes give $k_i-1\le k_j$.
Combining these two inequalities gives constructed EF1. This calculation
also covers prefixes lying inside a group of additional chores, during
which $A$ stays unchanged. The extended
allocation is therefore TEF1 throughout.

Together, the two directions show that the constructed instance admits a
TEF1 allocation if and only if the original instance does.

\subsection{Proofs of Theorem~\ref{thm:near-uniform} and Corollary~\ref{cor:uniform-limit}}

\begin{proof}[Proof of Theorem~\ref{thm:near-uniform}]
Fix $n\ge3$. Apply Lemma~\ref{lem:add-agents} to the three-agent instances
of Appendix~\ref{app:hardness}, then apply the transformation of
Appendix~\ref{app:near-uniform} to the resulting $n$-agent instances.
All costs are positive integers. Every cost before the latter transformation
is at most $S$, so choosing $K\ge\max\{S+1,\lceil S/\eta\rceil\}$ places every constructed cost in $[K,K+S]\subseteq[K,(1+\eta)K]$.
For the stronger restriction, choose instead $K=\max\{S+1,S\,2^{\widehat m}\}$,
where $\widehat m=n(m-n)+1$ is the number of constructed chores. Since every
constructed cost lies in $[K,K+S]$, we then have
$\max_{i,c}\widehat d_i(c)\le K+S\le(1+2^{-\widehat m})K\le(1+2^{-\widehat m})\min_{i,c}\widehat d_i(c)$.
The value $K$ has $O(\log S+\widehat m)$ bits, so the transformation takes
polynomial time and preserves the number of agents. By
Appendix~\ref{app:near-uniform}, it also preserves the answer to the existence
question, so NP-hardness follows for every fixed $n\ge3$. Membership in
NP follows from the same prefix-by-prefix check as before.
\end{proof}

\begin{proof}[Proof of Corollary~\ref{cor:uniform-limit}]
For $n=3$, apply the transformation to Table~\ref{tab:nine-chores}. Its largest row
total is $55$ and its largest entry is $10$. For every $K>55$, the
resulting order is $\widehat c_1,\widehat c_2,\widehat c_3$, followed by
$\widehat c_t,u,u$ for each $t=4,\dots,8$, and finally $\widehat c_9$.
There are nineteen chores, each of cost between $K$ and $K+10$.
No TEF1 allocation exists by Appendices~\ref{app:small-instance}
and~\ref{app:near-uniform}, giving
$(M_3,D_3,S_3)=(19,10,55)$.

For $n\ge4$, use the instance from
Appendix~\ref{app:general-nonexistence} with $\alpha=1$ and $h=4$. It has
$m_n=n+O(\log n)$ chores and no TEF1 allocation. Let $D_n$ be its largest
cost and $S_n$ its common row total. Both depend only on $n$. For every
integer $K>S_n$, the transformation produces an instance with
$M_n=n(m_n-n)+1=O(n\log n)$
chores, all costs in $[K,K+D_n]$, and, again by
Appendix~\ref{app:near-uniform}, no TEF1 allocation.

For either case and every $m>M_n$, append $m-M_n$ chores of cost $K$ to
every agent. Since fairness is required after every round, this does not
restore a TEF1 allocation. Division by $K$ places all costs in
$[1,1+D_n/K]$, which converges to $\{1\}$ while the matrix size stays fixed.
At the limiting all-ones matrix, assigning chores cyclically is TEF1 because
bundle sizes differ by at most one after every arrival.
\end{proof}

The instances obtained from the geometric construction also have the same
multiset of individual costs in every row: adding $K$ to each entry and
inserting columns of cost $K$ preserve this property. This also gives such
nineteen-chore instances for $n=3$, by using Table~\ref{tab:powers} with
$h=4$ rather than Table~\ref{tab:nine-chores}, although the constants
$D_3,S_3$ are then larger.

\section{Proof of Proposition~\ref{prop:cost-separation}}
\label{app:cost-separation}

\begin{proof}
We first treat $n=2$ and $m=5$.
Fix $0<\varepsilon\le1$ and $H>2$, and use Table~\ref{tab:cost-example}.
We work with the costs before normalization; all comparisons and cost
ratios are unchanged when the matrix is divided by $3H+1+\varepsilon$.

The sequence $1,2,1,1,2$ assigns every chore to a minimum-cost recipient,
so its social cost $3+2\varepsilon$ is the unconstrained optimum. In the
first two prefixes, every bundle contains at most one chore. After $c_3$,
agent $1$ has
own costs $\varepsilon,1$, so her residual is $\varepsilon\le1$, the
cost of agent $2$'s bundle. After $c_4$, her residual becomes
$1+\varepsilon$, while that comparison cost remains $1$. Agent $2$
still has a singleton, so this is the only inequality that requires a
factor greater than one.

After $c_5$ goes to agent $2$, agent $1$'s comparison cost becomes
$1+(3H-2)=3H-1\ge1+\varepsilon$, restoring her exact EF1 inequality.
Agent $2$ owns chores of costs $1,\varepsilon$. Her residual is
$\varepsilon$, since $\varepsilon\le1$, and she values the other
bundle at $3H$. Thus the final allocation is EF1, and the entire
sequence is $(1+\varepsilon)$-TEF1.

Now take any exact TEF1 allocation. Suppose none of $c_1,c_3,c_4$ goes
to agent $2$. All three then go to agent $1$. Since all costs are positive,
the first two chores cannot have the same recipient, so the second chore
goes to agent $2$. Agent $1$'s EF1 inequality after the fourth chore then
reads $1+\varepsilon\le1$, a contradiction. At least one of $c_1,c_3,c_4$
must therefore be allocated to agent $2$, at cost $H$.

For completeness, exact TEF1 is attainable through the sequence
$1,2,1,2,2$. Its first three prefixes were checked above. At the fourth
round, agent $2$ has residual $1$ and evaluates agent $1$'s bundle at
$2H$. At the fifth, this residual becomes $1+\varepsilon\le2H$.
Agent $1$ keeps residual $\varepsilon$, while her comparison bundle
only becomes more costly. Thus this allocation is TEF1 and has social
cost $H+2+2\varepsilon$.

The minimum exact TEF1 social cost is therefore at least $H$, while
the unconstrained optimum is $3+2\varepsilon$. Their ratio is at
least $H/(3+2\varepsilon)$ and tends to infinity with $H$.
Both row totals are $3H+1+\varepsilon$, so the same conclusions hold
with strictly positive normalized costs.

\paragraph{More agents and more chores.}
Now let $n\ge2$ and $m\ge n+3$, write $k=m-n-3$, and assume $H\ge3$.
Keep agents $1$ and $2$, add agents $3,\dots,n$, place $n-2$ new chores
before $c_1$, and append $k$ chores after $c_5$. Each new chore costs $1$
to every added agent and $H$ to agents $1$ and $2$. Each appended chore
costs $1$ to every agent. Each added agent has cost $H$ for each of
$c_1,\dots,c_4$ and cost $(n-3)(H-1)+\varepsilon$ for $c_5$. Every row
total is then $(n+1)H+1+\varepsilon+k$, so dividing all costs by this total
again normalizes the instance.

Give the new chores to distinct added agents, $c_1,\dots,c_5$ to agents
$1,2,1,1,2$ as before, and the appended chores to all agents in cyclic
order. Every chore goes to a minimum-cost agent, so this allocation attains
the unconstrained minimum social cost $n+1+2\varepsilon+k$. Up to round
$n+3$, each added agent owns a single chore, agents $1$ and $2$ compare
their bundles as before, and each of them evaluates the bundle of an added
agent at $H$, which exceeds her residual. Hence these prefixes are
$(1+\varepsilon)$-TEF1, and the prefix at round $n+3$ is EF1. Moreover, at
round $n+3$, every agent's cost for every other bundle exceeds her residual
by at least $1$; for instance, the difference is $H-1-\varepsilon\ge1$ for
agent $1$ and an added agent, and exactly $1$ for two added agents. Each
appended chore increases the residual of
its recipient by exactly $1$, because she already owns a chore of cost at
least $1$ to her, and increases every other agent's cost for the recipient's
bundle by $1$. Under the cyclic order, whenever an agent has received more
appended chores than another agent, she has received exactly one more. Each
difference therefore decreases by at most $1$ and stays nonnegative, so all
later prefixes are EF1. The sequence $1,2,1,2,2$ for $c_1,\dots,c_5$ is
also TEF1 with the added agents: the residuals of agents $1$ and $2$ never
exceed $1+\varepsilon<H$, their cost for each added agent's bundle.
After $c_5$, agents $1$ and $2$ have residuals $\varepsilon$ and
$1+\varepsilon$, respectively, and every added agent has residual zero.
Every comparison cost again exceeds the corresponding residual by at least
$1$, so the cyclic extension is TEF1 as well.

Finally, consider any TEF1 allocation. If it gives a new chore to agent $1$
or $2$, or one of $c_1,\dots,c_4$ to an added agent, its social cost is at
least $H$. Otherwise, the first $n$ chores have distinct recipients by
positivity, so the new chores go to the added agents and $c_1,c_2$ go to
agents $1$ and $2$ in some order, while $c_3,c_4$ also go to agents $1$ and
$2$. The two-agent argument above then shows that agent $2$ receives one of
$c_1,c_3,c_4$, at cost $H$. The ratio is therefore at least
$H/(n+1+2\varepsilon+k)$, which tends to infinity with $H$.
\end{proof}

\section{Proof of Theorem~\ref{thm:cost-hardness}}
\label{app:cost-hardness}

We prove Theorem~\ref{thm:cost-hardness} by fixing any polynomial
$\gamma(m)\ge1$ and showing that a polynomial-time
$\gamma(m)$-approximation would imply P$=$NP. Every constructed instance has
a TEF1 allocation that can be supplied explicitly. A solution to \partprob{}
is needed only to obtain a cheap allocation. Additional chores of cost $1$
to two agents make the supplied allocation possible, but these costs are too
small to change the partition encoded by a cheap allocation.

\subsection{The Construction}

Given positive integers
$a_1,\dots,a_r$, the constructed instance will have $m=3r+22$ chores.
Let $U=\sum_{j=1}^r a_j$, $L=r+4$, and $B=(L+1)U$.
Choose an integer
\begin{equation}\label{eq:large-cost}
 H>\gamma(3r+22)\,(68B+2L+1).
\end{equation}
For instance, $H=(\lceil\gamma(3r+22)\rceil+1)(68B+2L+1)$ suffices.
Begin with the nine-chore instance of Lemma~\ref{lem:unique-prefix}
with offset $s=L$. Its unique TEF1 allocation has bundle-cost matrix
\eqref{eq:prefix-matrix}, social cost $38B$, and own maxima and
residuals as in Lemma~\ref{lem:unique-prefix}.

Next specify $L$ chores $x_1,\dots,x_L$. For $j\le r$, the input chore
$x_j$ has cost vector $(2(L+1)a_j,\,H,\,2(L+1)a_j)$.
The remaining four cost vectors are $(6B,H,6B)$, $(8B,H,8B)$, $(6B,H,6B)$,
and $(8B,H,8B)$.
The total cost of these $L$ chores is $30B$ to each of $p,q$:
the input chores contribute $2B$ and the last four contribute $28B$.
Every one of them costs $H$ to $d$.

Immediately before each $x_j$, insert two chores of cost vector
$(1,H,1)$. There are $2L$ such chores. Their costs to $p$ and $q$ are
small, but assigning one to either of these agents increases $d$'s
cost for that bundle by $H$. This will allow an explicit allocation
in which $d$ receives all $x_j$.

We append one final chore to make all row totals equal. Before this
chore, the total costs of the entire instance to $p,d,q$ are
$73B+3L$, $31B+3LH$, and $85B+3L$, respectively. Write $D=31B+3LH$, and
give the final chore cost vector $(D-73B-3L+1,\,1,\,D-85B-3L+1)$.
All entries are positive. Indeed, the smaller of the first and third
entries is $D-85B-3L+1=3LH-54B-3L+1>0$, since $L\ge5$ and, by
\eqref{eq:large-cost}, $H>68B+2L+1$. After this final chore, every row total is
$D+1$. Divide every cost by $D+1$ to obtain normalized costs. We
continue the proof using the unnormalized numbers; the common division
preserves TEF1 and all multiplicative social-cost comparisons.

The construction has $9+3L+1=3r+22$ chores and, since $\gamma$ is fixed,
takes polynomial time. Unlike the existence reduction, it does not exclude
$d$ from all later chores. Instead, receiving any of the first $3L$ later
chores is expensive for $d$, and the proof distinguishes whether any
such assignment occurs.

\subsection{A Cheap TEF1 Allocation Implies a Partition}
\label{app:cheap-partition}

Suppose that a TEF1 allocation assigns none of the $L$ chores $x_j$
and none of the $2L$ inserted chores to $d$. At any prefix between the first
nine chores and the final chore, let $T_p,T_q$
be the total costs of the later chores given to $p,q$, respectively.
Consider the prefix just before that final chore. All chores after the first
nine up to this prefix have the same cost to $p$ and $q$, so
$T_p+T_q=30B+2L$.
The own maxima of $p,q$ remain $8B,10B$: input chores cost at most
$2B$, the last four $x_j$ cost at most $8B$, and inserted chores
cost $1$. Comparing with $d$'s unchanged bundle in
\eqref{eq:prefix-matrix}, with $s=L$, gives
$B+T_p\le14B+L$ and $8B+T_q\le25B+L$.
The two upper bounds sum to the total cost. They must therefore both
hold with equality, so $T_p=13B+L$ and $T_q=17B+L$.

Let $k$ be the number of inserted cost-$1$ chores given to $p$,
let $X$ be the cost of input chores $x_1,\dots,x_r$ given to $p$,
and let $Y$ be the cost of the last four $x_j$ given to $p$.
The first of these equalities gives $X+Y+k=13B+L$.
The numbers $X,Y,13B$ are all multiples of $L+1$. Hence $L-k$ is a
multiple of $L+1$. But $0\le k\le2L$, so $-L\le L-k\le L$;
the only possible multiple is zero. It follows that $k=L$ and
$X+Y=13B$.

Now $0\le X\le2B$, and $Y$ is a subset sum of
$6B,8B,6B,8B$. The same even-multiple argument as in
Appendix~\ref{app:hardness} gives $Y=12B$ and $X=B$.
Consequently, $2(L+1)\sum_{j\le r:\,x_j\in A_p}a_j=B=(L+1)U$, so those
input numbers sum to $U/2$.

Thus, when the \partprob{} instance has no solution, every TEF1
allocation must give at least one of the $L$ chores $x_j$ or one
of the $2L$ inserted chores to $d$. Its unnormalized social cost
is then at least $H$. The final chore does not affect this conclusion,
since the argument concerns the prefix before it.

\subsection{A Partition Gives a Cheap TEF1 Allocation}

Suppose a subset of the input numbers sums to $U/2$. Give its input
chores to $p$ and the remaining input chores to $q$, so each receives
input cost $B$. Give the last four $x_j$ to $p,q,p,q$.
Before every $x_j$, give the first inserted chore to $p$ and the
second to $q$. Give the final chore to $d$.
The social cost is $38B+30B+2L+1=68B+2L+1$.
We now check that the entire sequence is TEF1.

Before the final chore, agent $d$'s own bundle is unchanged and
her comparison costs only increase. For $p$ and $q$, the EF1
inequalities at every prefix between the first nine chores and the final
chore are now
\begin{equation}\label{eq:cost-four-constraints}
 T_p\le13B+L,\quad T_q\le17B+L,\quad
 T_p-T_q\le19B,\quad T_q-T_p\le4B.
\end{equation}
To see the first two, compare their residuals with $d$'s bundle in
\eqref{eq:prefix-matrix}, with $s=L$. The other two compare $p$ with $q$ and
$q$ with $p$, whose initial comparison costs did not change with
the offset $L$.

If inserted chores are temporarily omitted, $T_p$ and $T_q$ follow the
allocation in Appendix~\ref{app:hardness}: while the input chores arrive,
both are at most $B$, and after the last four $x_j$ the pairs $(T_p,T_q)$ are
$(7B,B)$, $(7B,9B)$, $(13B,9B)$, $(13B,17B)$. Each agent receives at most
$L$ inserted chores, so the first two inequalities in
\eqref{eq:cost-four-constraints} hold throughout.
After each complete inserted pair, its contributions to $T_p$ and $T_q$
are equal. Between its first and second chores, the difference
$T_p-T_q$ is larger by one. Thus $T_q-T_p$ never exceeds the bound
$4B$ from the allocation without the pairs, and $T_p-T_q$ is at
most $6B+1\le19B$. This checks the other two inequalities at every
intermediate prefix, including those inside a pair.

Giving the final chore to $d$, whose cost for it is $1$, increases her
residual from $6B$ to $6B+1$. Each other agent has already received
$L$ inserted chores of cost $H$ to $d$, so both comparison bundles
cost more than $6B+1$. Her EF1 inequalities hold. The comparison
bundle of each of $p,q$ only becomes more costly, so their
inequalities continue to hold as well. Hence the cheap allocation is TEF1.

\subsection{The Supplied TEF1 Allocation}

We next construct the allocation supplied with every instance.
Assign the first nine chores as in Lemma~\ref{lem:unique-prefix}. Before
each $x_j$, give the
two inserted chores to $p$ and $q$, in that order, and give every
$x_j$ to $d$. Give the final chore to $d$ as well. This allocation
can be specified without solving \partprob{}.

After the first nine chores, agents $p$ and $q$ receive only chores that
cost $1$ to them, so their maxima remain $8B$ and $10B$. With $T_p,T_q$ defined as before, every prefix therefore
has $0\le T_p,T_q\le L$ and $T_p-T_q\in\{0,1\}$, so all four inequalities
in \eqref{eq:cost-four-constraints} hold. These inequalities still suffice
for $p$ and $q$: the last two do not involve $d$, and giving chores to $d$
only increases the comparison costs in the first two.

Before $d$ receives her first $x_j$, her residual is still $6B$ and her
comparison costs have only increased. After receiving $k\ge1$ such chores,
her own cost is $11B+kH$ and her largest own cost is $H$, because
$H>5B$. Her residual is therefore $11B+(k-1)H$.
At that moment, each of $p,q$ has received exactly $k$ inserted
chores of cost $H$ to $d$. Her two comparison costs are
$6B+kH$ and $14B+kH$. The smaller of these is at least
her residual precisely when $H\ge5B$, which holds by
\eqref{eq:large-cost}. Between two $x_j$, inserting the next
pair only increases her comparison costs.

The final chore adds $1$ to $d$'s residual, leaving
$11B+(L-1)H+1$. This is at most $6B+LH$ because $H\ge5B+1$.
The other agents continue to satisfy EF1 as their own bundles do
not change. Hence the supplied allocation is TEF1 at every round.

\subsection{Completing the Proof}

\begin{proof}[Completion of the proof of Theorem~\ref{thm:cost-hardness}]
After normalization, the instance constructed from a yes-instance of
\partprob{} has a TEF1 allocation with social cost at most
$(68B+2L+1)/(D+1)$, while every TEF1 allocation of the instance constructed
from a no-instance has social cost at least $H/(D+1)$. Every constructed
instance comes with the explicit allocation just shown to be TEF1, as
required in the theorem.

Suppose that a polynomial-time algorithm returned, on these instances, a
TEF1 allocation whose social cost is at most $\gamma(m)$ times the minimum
social cost of a TEF1 allocation. In a yes-instance, its output would have
social cost at most $\gamma(3r+22)(68B+2L+1)/(D+1)<H/(D+1)$,
whereas in a no-instance its TEF1 output would have cost at least
$H/(D+1)$. Comparing its output with this threshold would decide
\partprob{} in polynomial time. Therefore such an algorithm would imply
P$=$NP.
\end{proof}

\subsection{More Than Three Agents}

Finally, fix $n\ge4$. We extend the construction by adding $n-3$ agents and
$n-3$ new chores that only the added agents can perform inexpensively.
Choose $H$ so that, in addition to \eqref{eq:large-cost},
$H>\gamma(3r+19+n)\,(68B+2L+n-2)$.
Place the new chores before all chores of the three-agent instance. Each
new chore costs $1$ to every added agent and $D+1$ to each of $p,d,q$. Every
other chore costs $H$ to every added agent, except the final chore, which
costs $(n-2)(D+1)-(n-3)-(3r+21)H$ to every added agent. This cost is positive: since $n\ge4$ and
$D>3LH=(3r+12)H$, it is at least $2(D+1)-1-(3r+21)H>(3r+3)H$. All row
totals are now $(n-2)(D+1)$, and the instance has $3r+19+n$ chores.

Every TEF1 allocation of the three-agent instance extends to a TEF1
allocation of the new instance, with social cost larger by $n-3$, by giving
the new chores to distinct added agents. Indeed, each added agent then owns
a single chore throughout, and each of $p,d,q$ evaluates such a bundle at
$D+1$, which exceeds her residual because $D+1$ is her total cost in the
three-agent instance. In particular, the supplied allocation extends, and a
yes-instance of \partprob{} yields a TEF1 allocation with unnormalized social
cost $68B+2L+n-2$.

Conversely, consider any TEF1 allocation of the new instance with
unnormalized social cost less than $H$. Since $D+1>H$, it gives no new chore
to $p$, $d$, or $q$, and it gives no chore other than the final one to an
added agent. Removing the new chores and the added agents therefore leaves
a TEF1 allocation of the three-agent instance before its final chore, with
the first nine recipients fixed by Lemma~\ref{lem:unique-prefix}.
Moreover, $d$ receives none of the chores that cost $H$ to her, so the
argument in Appendix~\ref{app:cheap-partition} yields a solution to
\partprob{}. Hence, for a no-instance, every TEF1 allocation has unnormalized
social cost at least $H$.
Repeating the completion of the proof above with $68B+2L+n-2$, $3r+19+n$,
and $(n-2)(D+1)$ in place of $68B+2L+1$, $3r+22$, and $D+1$ proves
Theorem~\ref{thm:cost-hardness} for $n$ agents.

\section{Proof of Theorem~\ref{thm:fptas}}
\label{app:fptas}

We give the full algorithms and proof of Theorem~\ref{thm:fptas}.
The two dynamic programs use the same description of a partial allocation
and the same geometric grouping. They differ only in how they choose an
allocation within a group. The first minimizes the attained fairness
factor; the second minimizes social cost subject to a threshold on that
factor that increases slightly with each round. We analyze them separately before combining
their guarantees.

We assume $m\ge1$, since the empty instance has factor $1$ and social
cost zero. For the numerical analysis we take costs to be integers,
clearing denominators by a common scaling when necessary. Such a scaling
preserves every fairness ratio and multiplies every social cost by the
same amount, so neither guarantee is affected.

\subsection{Exact Information Associated with a Partial Allocation}
\label{app:dp-state}

For a partial allocation $A$, retain the vector of exact maxima
$M_i=M_i(A)$ and the $n^2$ nonnegative coordinates $V_{ii}=R_i(A)$ and
$V_{ij}=d_i(A_j)$ for $i\ne j$.
The diagonal contains residual costs, not the costs of own bundles. Let $G$ be the
temporal EF1 factor of this partial allocation, defined as in
\eqref{eq:factor} with $t$ ranging over the rounds so far, and let $W$ be its
social cost so far.
A state consists of $M,V,G,W$, together with a pointer to its preceding
state and the recipient of its last chore. Every state describes an
actual allocation, and every stored quantity is exact. The grouping
introduced below does not replace these quantities by rounded values.

If the next chore $c$ is assigned to agent $i$, the coordinates change
as follows:
\begin{equation}\label{eq:dp-update}
\begin{aligned}
 V'_{ii}&=V_{ii}+\min\{M_i,d_i(c)\},
 &M'_i&=\max\{M_i,d_i(c)\},\\
 V'_{ji}&=V_{ji}+d_j(c) &&(j\ne i).
\end{aligned}
\end{equation}
Every other coordinate and maximum stays unchanged, and
$W'=W+d_i(c)$. Formula~\eqref{eq:dp-update} follows from
\eqref{eq:residual-update} for the diagonal and additivity for the
other entries. In particular, two states with the same exact maximum
vector receive identical nonnegative increments to $V$ when extended
using the same recipient.

Only the recipient's ratios can increase. Consequently the exact factor $G$
is updated according to
\begin{equation}\label{eq:dp-factor-update}
 G'=\max\left\{G,\max_{j\ne i}\frac{V'_{ii}}{V'_{ij}}\right\}.
\end{equation}
The quotient conventions are those in \eqref{eq:factor}.
If $G'=+\infty$, the state can be discarded: later assignments cannot
alter a fairness ratio at an earlier prefix. Such a state cannot be
part of any allocation with finite temporal EF1 factor. Since
$\alpha^*$ is finite, no prefix of an allocation attaining $\alpha^*$
is discarded in this way.

\subsection{Grouping States and Controlling the Error}
\label{app:grouping}

Fix an auxiliary accuracy $0<\rho\le1$, and set $b=1+\rho/(4m)$ and
$S=\max\{1,\max_i d_i(C)\}$.
Every coordinate $V_{ij}$ lies in $[0,S]$.
We treat zero separately, and assign a positive coordinate $x$ to the
unique interval $[b^k,b^{k+1})$ containing $x$.
Here $k\ge0$ because positive coordinates are positive integers.
Two states are in the same group when their exact maximum vectors
agree and every coordinate of $V$ belongs to the same interval,
with zero matching only zero. The values $G$ and $W$ are not part
of the group description; they determine which allocation is kept.

There are two estimates behind the proof. First, for $\lambda\ge1$,
nonnegative $x,y,z$ satisfy
\begin{equation}\label{eq:common-increments}
 \lambda^{-1}x\le y\le\lambda x
 \quad\Longrightarrow\quad
 \lambda^{-1}(x+z)\le y+z\le\lambda(x+z).
\end{equation}
For the lower bound, use $z\ge\lambda^{-1}z$; for the upper bound,
use $z\le\lambda z$. Thus adding the same nonnegative amount preserves
a multiplicative comparison. Second, coordinates in the same
positive interval differ by a factor of less than $b$, while
a zero coordinate is grouped only with zero. Therefore, replacing a state
by another state in the same group changes each coordinate by a factor of
less than $b$.

For a reference allocation $\pi$, write $M_i^\pi,V_{ij}^\pi$ for the
maxima and coordinates of its prefix at the current round. In the inductive
proofs below, we use these two estimates to find a retained state with the
same exact maxima as the reference prefix and whose coordinates satisfy
\begin{equation}\label{eq:tracking-coordinates}
 b^{-t}V^{\pi}_{ij}\le V_{ij}\le b^tV^{\pi}_{ij}
 \quad\text{for all }i,j.
\end{equation}
Consequently, for $i\ne j$ with finite $V^\pi_{ii}/V^\pi_{ij}$, we have
$V_{ii}/V_{ij}\le b^{2t}\,V^\pi_{ii}/V^\pi_{ij}$.
Zero coordinates are handled by the same inequalities: if the
reference coordinate is zero, the retained coordinate must also
be zero. If a reference comparison cost is positive, the retained
comparison cost is positive as well. Hence a finite ratio for $\pi$
cannot correspond to an infinite ratio for the retained state.

These estimates by themselves do not specify which state should
be kept. Keeping a state merely because its coordinates are close
could select a partial allocation that had a poor fairness ratio
much earlier. The two programs below therefore select states using
the exact factor $G$, and we prove bounds on $G$ together with
\eqref{eq:tracking-coordinates}.

\subsection{The Two Dynamic Programs}
\label{app:dp-algorithms}

Algorithms~\ref{alg:fairness-dp} and~\ref{alg:cost-dp} use the exact
extension in \eqref{eq:dp-update}--\eqref{eq:dp-factor-update}.
The empty state has $M=0,V=0,G=1,W=0$. Whenever a state is retained,
its predecessor pointer refers to the retained state from the
previous round from which it was extended. Ties may be broken
arbitrarily and consistently.

\begin{algorithm}[ht]
\caption{Approximate the minimum temporal EF1 factor}
\label{alg:fairness-dp}
\begin{algorithmic}[1]
\Require An ordered instance and $0<\rho\le1$
\State Form the groups using $b=1+\rho/(4m)$.
\State $\mathcal S\gets\{\text{empty state}\}$.
\For{$t=1,\dots,m$}
  \State $\mathcal T\gets\varnothing$.
  \For{each state in $\mathcal S$ and each recipient $i\in N$}
    \State Extend the state by assigning $c_t$ to $i$ using
      \eqref{eq:dp-update}--\eqref{eq:dp-factor-update}.
    \If{the resulting $G$ is finite}
      \State In its group in $\mathcal T$, keep the state with smallest $G$.
    \EndIf
  \EndFor
  \State $\mathcal S\gets\mathcal T$.
\EndFor
\State \Return the allocation from a state in $\mathcal S$ with smallest $G$.
\end{algorithmic}
\end{algorithm}

\begin{algorithm}[ht]
\caption{Minimize cost with a small relaxation of a target factor}
\label{alg:cost-dp}
\begin{algorithmic}[1]
\Require An ordered instance, target $\alpha\ge1$, and $0<\rho\le1$
\State Form the groups using $b=1+\rho/(4m)$.
\State $\mathcal S\gets\{\text{empty state}\}$.
\For{$t=1,\dots,m$}
  \State $\mathcal T\gets\varnothing$.
  \For{each state in $\mathcal S$ and each recipient $i\in N$}
    \State Extend the state by assigning $c_t$ to $i$ using
      \eqref{eq:dp-update}--\eqref{eq:dp-factor-update}.
    \If{the resulting $G\le b^{2t}\alpha$}
      \State In its group in $\mathcal T$, keep the state with smallest $W$.
    \EndIf
  \EndFor
  \State $\mathcal S\gets\mathcal T$.
\EndFor
\If{$\mathcal S=\varnothing$}
  \State \Return no allocation.
\Else
  \State \Return the allocation from a state in $\mathcal S$ with smallest $W$.
\EndIf
\end{algorithmic}
\end{algorithm}

The second algorithm is guaranteed to return an allocation whenever
the target $\alpha$ is attainable. It can also return an allocation
when the exact target is not attainable, since the permitted
factor is larger. Thus its output does not decide whether an
$\alpha$-TEF1 allocation exists.

\subsection{Approximating the Optimal Factor}
\label{app:fairness-induction}

Fix an allocation $\pi^*$ attaining $\alpha^*$. We prove the
following inductive claim for Algorithm~\ref{alg:fairness-dp}:
after round $t$, some retained state satisfies
\begin{equation}\label{eq:fairness-invariant}
 \begin{split}
 M_i&=M_i^{\pi^*}\quad\text{for all }i,\\
 b^{-t}V_{ij}^{\pi^*}&\le V_{ij}\le b^tV_{ij}^{\pi^*}
       \quad\text{for all }i,j,\\
 G&\le b^{2t}\alpha^*.
 \end{split}
\end{equation}
The claim holds at round zero for the empty state.

Suppose it holds after round $t-1$. Extend the state given by the
induction hypothesis using the recipient of $c_t$ in $\pi^*$, and call
the result the candidate. Because the exact maximum vectors of this
state and of the prefix of $\pi^*$ agree before the extension,
\eqref{eq:dp-update} adds the same nonnegative amounts to their
corresponding coordinates, and the maximum vectors still agree
afterwards.
By \eqref{eq:common-increments}, the candidate's coordinates
remain within a factor $b^{t-1}$ of the reference coordinates.
Every new recipient ratio is therefore at most
$b^{2(t-1)}\alpha^*$, using the zero-coordinate argument
following \eqref{eq:tracking-coordinates} when needed. The
value of $G$ before the extension has the same upper
bound by the induction hypothesis. Equation~\eqref{eq:dp-factor-update}
then gives $G\le b^{2(t-1)}\alpha^*$ for this candidate, in particular a finite value.

The state finally kept in its group has no larger $G$, because
Algorithm~\ref{alg:fairness-dp} minimizes $G$ within each group.
It has the same maxima and differs from the candidate in each
coordinate by at most a factor $b$. Its coordinates are therefore
within a factor $b^t$ of those of $\pi^*$, and its value of $G$
is at most $b^{2t}\alpha^*$. This completes the
induction.

Let $\overline\alpha$ be the actual factor of the allocation
returned by Algorithm~\ref{alg:fairness-dp}. The final state
with smallest $G$ is at least as good as the one in
\eqref{eq:fairness-invariant}. Since it describes an actual
allocation, we obtain
\begin{equation}\label{eq:fairness-first-bound}
 \alpha^*\le\overline\alpha
 \le b^{2m}\alpha^*
 \le e^{\rho/2}\alpha^*
 \le(1+\rho)\alpha^*.
\end{equation}
Here $(1+x)\le e^x$ gives the penultimate inequality. For the
last inequality, $\log(1+\rho)\ge\rho/2$ on $[0,1]$:
the derivative of $\log(1+\rho)-\rho/2$ is nonnegative on that
interval, and the difference is zero at $\rho=0$.

\subsection{Preserving Cost Relative to an Exact Target}
\label{app:cost-induction}

Now fix a target $\alpha\ge1$ and any allocation $\pi$ with
$\alpha(\pi)\le\alpha$. Let $W^\pi_t$ be its social cost through
round $t$. For Algorithm~\ref{alg:cost-dp}, we prove that after
every round $t$ some retained state satisfies
\begin{equation}\label{eq:cost-invariant}
 \begin{split}
 M_i&=M_i^\pi\quad\text{for all }i,\\
 b^{-t}V_{ij}^\pi&\le V_{ij}\le b^tV_{ij}^\pi
        \quad\text{for all }i,j,\\
 G&\le b^{2t}\alpha,\quad W\le W^\pi_t.
 \end{split}
\end{equation}
The claim holds at round zero for the empty state.

Assume the claim after round $t-1$, and extend the state given by
the induction hypothesis using the recipient of $c_t$ in $\pi$ to obtain
a candidate. Equal exact maxima again give identical coordinate increments. The candidate is
within a factor $b^{t-1}$ of the reference in every coordinate,
and its new ratios are at most $b^{2(t-1)}\alpha$.
The value of $G$ before the extension obeys the same bound. Hence the
candidate satisfies the round-$t$ threshold $G\le b^{2t}\alpha$ and is considered
in its group. Also, the candidate and the reference add the same cost
$d_{\pi(t)}(c_t)$, so the candidate still satisfies
$W\le W^\pi_t$.

The retained state in this group has no larger social cost,
because Algorithm~\ref{alg:cost-dp} keeps the cheapest state.
It may have a larger value of $G$ than the candidate;
this does not affect the argument, because every state considered
in the group satisfies the threshold $G\le b^{2t}\alpha$.
Its maxima agree exactly and its coordinates are within another
factor $b$ of the candidate. All parts of
\eqref{eq:cost-invariant} therefore hold after the selection.
This completes the induction.

Choose $\pi$ to be a minimum-cost allocation whose factor is at
most $\alpha$. Such an allocation exists whenever the target
is attainable. Taking the cheapest final state and using
\eqref{eq:cost-invariant} and the estimate from
\eqref{eq:fairness-first-bound}, Algorithm~\ref{alg:cost-dp}
returns an allocation $\pi_\alpha$ with
\begin{equation}\label{eq:target-bound}
 \begin{split}
 \alpha(\pi_\alpha)&\le b^{2m}\alpha\le(1+\rho)\alpha,\\
 \mathrm{sc}(\pi_\alpha)&\le
 \min_{\pi:\,\alpha(\pi)\le\alpha}\mathrm{sc}(\pi).
 \end{split}
\end{equation}
The social-cost inequality has no approximation term. Its
validity comes from selecting the minimum exact cost in every
group, rather than rounding cost or bounding it through other
coordinates.

\subsection{Combining the Guarantees}

\begin{proof}[Proof of the guarantees in Theorem~\ref{thm:fptas}]
For the target-factor statement, use
Algorithm~\ref{alg:cost-dp} with $\rho=\varepsilon$ and apply
\eqref{eq:target-bound}.

For the simultaneous optimal-factor and social-cost guarantees,
set $\rho=\varepsilon/3$. First run
Algorithm~\ref{alg:fairness-dp}, and let $\overline\alpha$ be
its exact attained factor. Then run
Algorithm~\ref{alg:cost-dp} with target $\overline\alpha$ and
the same $\rho$. This target is attainable because the first
algorithm has just produced an allocation attaining it.
Moreover, $\overline\alpha\ge\alpha^*$, so every allocation
attaining $\alpha^*$, including one that defines
$\mathrm{sc}^*$, has factor at most $\overline\alpha$.
By \eqref{eq:target-bound}, the output $\widehat\pi$ therefore satisfies $\mathrm{sc}(\widehat\pi)\le\mathrm{sc}^*$.
For its fairness factor, combine \eqref{eq:fairness-first-bound}
and \eqref{eq:target-bound} to obtain
$\alpha(\widehat\pi)\le(1+\rho)\overline\alpha\le(1+\rho)^2\alpha^*\le(1+\varepsilon)\alpha^*$.
The last inequality follows from
$2\varepsilon/3+\varepsilon^2/9\le\varepsilon$ for
$0<\varepsilon\le1$. The remaining running-time claim is
proved below.
\end{proof}

\subsection{Running Time and Reconstruction}
\label{app:running-time}

The number of interval choices for one coordinate, including
zero, is at most $Q=2+\lceil\log_b S\rceil$, which is
${O\bigl(m(1+\log S)/\rho\bigr)}$.
To obtain the bound, use
$\log(1+\rho/(4m))\ge\rho/(8m)$, since
$0<\rho/(4m)\le1/4$.
Every own maximum $M_i$ is either zero or the cost of one of the
$m$ chores. Consequently there are at most $(m+1)^n$ exact
maximum vectors. Each round retains at most $(m+1)^nQ^{n^2}$ states. There are $m$ rounds and $n$ possible recipient extensions
per retained state. Forming the coordinates, computing a group,
and choosing its best state can all be done in polynomial time.
For fixed $n$, this bound is polynomial
in the input length and $1/\rho$.

All arithmetic takes polynomial time. The coordinates and maxima are
integers at most $S$, and social cost is at most $nS$. Each finite value of
$G$ is either $1$ or an attained ratio of two coordinate values, as is the
target $\overline\alpha$ passed from the first program to the second.
Interval membership can be determined by comparing with successive exact
powers of $b$, so no approximate logarithm is needed. These powers and the
thresholds $b^{2t}\alpha$ can be computed and compared in polynomial time.
Common scaling of the original costs, when needed, keeps $\log S$
polynomial in the input length.

Finally, keep the retained states of earlier rounds and their
predecessor pointers. At most $m(m+1)^nQ^{n^2}$ such states are needed. Following the
pointers from the chosen final state recovers one consistent
recipient for each chore. Because all states describe actual
allocations and all values of $G$ were computed exactly,
the reconstructed allocation has precisely the fairness factor
and social cost analyzed above. For fixed $n$, both algorithms therefore
run in time polynomial in the input length and $1/\rho$, which completes the
proof of Theorem~\ref{thm:fptas}.

\section{Additional Related Work}
\label{app:related}

This appendix complements Section~\ref{sec:related} with a few further
connections.

\paragraph{Nonexistence and complexity of fair allocations.}
Maximin share allocations need not exist for
goods \citep{KurokawaProcacciaWang2018} or for chores~\citep{AzizEtAl2017};
\citet{FeigeSapirTauber2021} gave nonexistence instances with
three agents and nine goods or nine chores, and showed that nine goods are
necessary for three agents. For chores, EFX allocations need not exist for
three agents with superadditive
costs~\citep{ChristoforidisSantorinaios2024}, and \citet{HeTao2026}
recently showed that EFX can fail with additive costs for four or more
agents. Our Theorems~\ref{thm:unbounded} and~\ref{thm:near-uniform} apply to every
fixed $n\ge3$, with additive costs and EF1, whose static version is always
satisfiable. Likewise, deciding whether an
envy-free allocation exists is NP-complete already for two agents with
identical additive valuations, through \partprob{}~\citep{LiptonEtAl2004};
since static EF1 allocations always exist, the hardness in
Theorem~\ref{thm:near-uniform} stems from the temporal requirement.

\paragraph{Ordered items and dynamic programs.}
\citet{HohneVanStee2021} studied the price of fairness
for indivisible chores allocated in contiguous blocks along a line; their
model also involves an order of the items, but it constrains only the final
allocation. Exact dynamic programs over partial allocations underlie the
pseudo-polynomial algorithms for welfare under EF1 with a fixed number of
agents~\citep{AzizEtAl2023EJOR}, for TEF1 existence~\citep{ChoiLiTeh2026}, and
for cumulative envy~\citep{GoldbergRobinsonTeh2026}. The distinctive features
of our scheme are that it keeps the largest own-chore costs exact while
grouping residuals, that it stores the exact fairness factor over all
earlier prefixes, and that it selects states so that social cost incurs no
loss. For constrained problems, one may approximate the objective while
keeping the constraint exact, or relax the constraint while matching the
optimal objective, as for generalized assignment~\citep{ShmoysTardos1993}. For
TEF1, Theorem~\ref{thm:cost-hardness} rules out the first option within any
polynomial factor, whereas Theorem~\ref{thm:fptas} achieves the second.

\paragraph{Temporal voting.}
In temporal (or perpetual) voting, a fixed group of voters makes a
collective decision in each of several rounds~\citep{Lackner2020};
\citet{ElkindObraztsovaTeh2024} propose a unified framework for these models.
Proportionality axioms for this setting have been strengthened beyond extended
justified representation to notions that remain
satisfiable~\citep{PhillipsEtAl2026}. Verifying such axioms is harder than in
multiwinner voting~\citep{ElkindEtAl2025Verifying}, although a strengthening of
full justified representation can be both computed and verified in
polynomial time~\citep{Teh2026FJR}, and ranked preferences have also been
considered~\citep{HazonSchmerlerTeh2026}. Maximizing utilitarian welfare is
easy, whereas optimizing egalitarian welfare is intractable even in restricted
cases, including for public chores~\citep{ElkindNeohTeh2024,ElkindNeohTeh2025},
and enforcing proportionality can incur a welfare loss that grows sublinearly
with the number of voters or rounds~\citep{Teh2026Price}. Related models
assign projects to time slots~\citep{ElkindKraiczyTeh2022} or seek a
second-stage committee that overlaps as much as possible with the
first~\citep{ZechEtAl2024}. As in our model, earlier decisions are typically
not revised. Each decision, however, is a public outcome shared by all voters,
whereas each chore here is performed by a single agent.

\paragraph{Online fair division.}
Against an oblivious adversary, for every $\alpha\in(0,1]$, every randomized
algorithm for online fair division of goods has an input on which it returns
an $\alpha$-EF1 allocation with probability at most
$e^{-\Omega(n)}$~\citep{CohenTehWooldridge2026}. Under budget constraints, no
deterministic online algorithm guarantees any fixed approximation to feasible
envy-freeness without further structure~\citep{CohenEtAl2026Budget}. In online
house allocation, envy-freeability with subsidies can always be maintained
using bounded recourse, although reassignment chains of length linear in the
number of agents are unavoidable in the worst case~\citep{TehEtAl2026House}.
Under upgrade recourse, the best maximin share guarantees differ sharply
between goods and chores~\citep{HuangTeh2026Chores,HuangTeh2026Goods}. These
models limit knowledge of future items, whereas ours provides it in full but
allows no revisions.

\paragraph{Fairness at every prefix.}
In online fair division, approximate fairness at every round has been
studied with an unknown horizon~\citep{KahanaSegalHaleviHazon2026}, and
competitive ratios have been measured with respect to the best fairness
level achievable offline~\citep{ChenTan2026}. Requirements that hold at every
prefix of a fixed sequence also arise in the chairman assignment
problem~\citep{Tijdeman1980}, where every prefix can be kept within less than
one assignment of proportional shares. For the envy-based requirement of EF1
with heterogeneous additive costs, Theorem~\ref{thm:unbounded} shows that no
such uniform guarantee is possible, even approximately and with complete
knowledge of the sequence.

\end{document}